\documentclass[a4paper,fleqn,usenatbib]{mnras}

\usepackage{newtxtext,newtxmath}
\usepackage[T1]{fontenc}
\usepackage{ae,aecompl}

\usepackage{graphicx}	
\usepackage{amsmath}	
\usepackage{amssymb}	
\usepackage[compatibility=false]{caption}
\usepackage{subcaption}
\usepackage{xspace}
\usepackage{hyperref}
\newcommand{\chromos}{\textsc{chromos}\xspace}

\title[Comparison of accreting NS and BH variability]{A model-independent comparison of the variability of accreting neutron stars and black holes}

\author[D.W. Gardenier and P. Uttley]{
D.W. Gardenier$^{1,2}$\thanks{E-mail: gardenier@uva.nl}
and P. Uttley$^{2}$
\\
$^{1}$ASTRON, the Netherlands Institute for Radio Astronomy, Postbus 2, 7990 AA, Dwingeloo, The Netherlands \\
$^{2}$Anton Pannekoek Institute for Astronomy, University of Amsterdam, Science Park 904, 1098 XH Amsterdam, The Netherlands
}

\date{Accepted XXX. Received YYY; in original form ZZZ}

\pubyear{2018}

\begin{document}
\label{firstpage}
\pagerange{\pageref{firstpage}--\pageref{lastpage}}
\maketitle

\begin{abstract}
We use Rossi X-ray Timing Explorer observations to conduct a population study of the timing properties of accretion-powered neutron star Low Mass X-ray Binaries (LMXBs), following a model-independent approach that was originally applied to black hole systems. The ratios of integrated power in four equally spaced Fourier frequency bands allow power spectral shapes to be parameterised with two `power colour' values, providing a simple way of tracking the evolution in timing properties across observations. We show that neutron star LMXBs follow a remarkably similar power spectral evolution to black hole LMXBs, confirming that the broadband noise variability seen in both types of system has a common origin in the accretion flow.  Both neutron stars and black holes follow a similar clear track in the power colour-colour diagram as they transition from the hard through soft states.  Quantifying the position on this oval track using a single parameter, the power-spectral `hue', we find that the transition in X-ray spectral shape occurs at the same hue for both neutron star and black hole systems.  The power colours of Z sources map on to those of soft state black holes, while those of atoll sources correspond to all spectral states.  There is no clear dependence of power colour on neutron star spin, or on whether the neutron star is clearly magnetised (determined by ms X-ray pulsations).
\end{abstract}

\begin{keywords}
 accretion, accretion discs -- black hole physics -- X-rays: binaries -- X-rays: individual: Aquila X-1
\end{keywords}



\section{Introduction}
Accreting compact objects have long been known to show related continuum variability properties, both across the mass scale \citep{Uttleyetal2005,McHardyetal2006} and across different compact object types \citep{WijnandsvanderKlis1999,Mauche2002,Uttley2004,Scaringietal2012}, which are thought to indicate a common origin of the variability in a turbulent accretion flow (e.g. \citealt{Lyubarskii1997}).  The similarities between the variations of neutron star and black hole low mass X-ray binaries (LMXBs) in particular, allow us to isolate which features are common to accretion flows in general, in objects with relatively similar mass (as compared to AGN) and compactness (as compared to accreting white dwarfs).   The differences in variability properties between neutron star and black hole LMXBs allow us to test for additional effects due to the presence of a solid surface versus event horizon and in some cases, a strong magnetic field.

Observations with the Rossi X-ray Timing Explorer ({\it RXTE}) demonstrated that those systems with a weak magnetic field, but which could be unambiguously identified as neutron stars through the presence of thermonuclear X-ray bursts, show systematically larger variability power than black holes at Fourier frequencies $\gtrsim 10$--$50$~Hz \citep{SunyaevRevnivtsev2000}.  The larger high-frequency variability amplitudes seen in neutron stars can be explained by the emission from the boundary layer at the neutron star surface, which can act as an amplifier for high-frequency signals produced in the innermost accretion flow to appear in the observed light curves.  The boundary layer may also play a role in producing the kHz quasi-periodic oscillations seen in accreting neutron stars, which do not have an obvious counterpart in black hole systems \citep{Mottaetal2017}, even allowing for the $\sim$few-hundred Hz high-frequency QPOs in black hole LMXBs, which appear rarely, during luminous transitional states.

If the highest frequencies show clear differences between neutron star and black hole systems, the lower-frequency variability is expected to show strong similarities, since it probes time-scales corresponding to larger scales in the accretion flow, further from the boundary layer or event horizon and thus similar (in terms of dynamics and the fraction of radiative power liberated) in both kinds of system.  Indeed, different characteristic features which appear as low-frequency breaks, bumps or sharper QPOs in the broadband power spectrum, do appear to follow the same frequency-frequency correlations in neutron star and black hole systems \citep{WijnandsvanderKlis1999,Psaltisetal1999,KleinWoltvanderKlis2008}.

In a given source, the frequencies and amplitudes of the power-spectral components also correlate strongly with changes in the X-ray spectrum and corresponding evolution through different spectral states \citep{HasingervanderKlis1989,Homanetal2001,Bellonietal2005}, indicating that the timing properties are closely linked to the structure of the inner emitting region.   Black hole LMXB X-ray spectra are commonly classified as hard, soft or intermediate state depending respectively on whether the luminosity is dominated by hard power-law emission, soft disk blackbody emission or some combination of disk and steep power-law emission.  The path of an outbursting black hole LMXB through the various states can be easily determined via a `hardness-intensity diagram' (HID, e.g. \citealt{Miyamotoetal1991,Homanetal2001,Bellonietal2005}) using the count rate ratio between two bands as a proxy for spectral shape.  Broadly speaking, frequencies in the power-spectrum are seen to increase through the hard-to-soft state transition with QPOs becoming relatively stronger compared to the broadband noise, while the overall rms amplitude (integrated over a broad frequency range) drops.   The timing properties in particular are linked to further distinctions, e.g. between the so-called hard-intermediate and soft-intermediate states \citep{HomanBelloni2005,Bellonietal2005}, which show respectively strong low-frequency (0.1--10~Hz) QPOs (type C, \citealt{Remillardetal2002}) with moderately strong broadband noise, versus generally weaker and less-coherent QPOs (type B or A, \citealt{Wijnandsetal1999}) superimposed on much weaker broadband noise (see also \citealt{Casellaetal2005} for a summary of low-frequency QPO classes).

Neutron star LMXBs also show evidence for similar states to black hole systems in their spectra and HIDs \citep{DoneGierlinski2003,Gladstoneetal2007}, but direct spectral comparison is not straightforward due to the additional emission associated with the boundary layer.  Neutron star LMXBs are also commonly classified into so-called `atoll' or `Z' sources based on the shapes they trace in colour-colour diagrams (CCDs, \citealt{HasingervanderKlis1989}), plots of flux ratios which are obtained using two pairs of soft and hard bands, with further sub-classifications made according to the detailed shape in the CCD.  Systems are usually uniquely classified as an atoll or Z source based on their CCD shape, however ambiguities remain close to the boundaries of the classes, where they appear to be similar to one another \citep{Munoetal2002,GierlinskiDone2002}.
The timing properties of neutron star LMXBs are also well correlated with their spectral shapes within the different classes, with atoll sources showing behaviour which is qualitatively equivalent to the full range of hard to soft states in black holes, while Z sources show timing behaviour similar to the intermediate to soft-state range \citep{vanderKlis2006}.

Given the complexity of comparing spectral shapes between neutron star and black hole LMXBs, and the evident similarities in their timing properties, it is useful to consider timing-based approaches for the comparison of neutron star and black hole evolution through the different states, which can give insight about the corresponding changes in inner region structure and the possible effects of the boundary layer and magnetic field.  Comparisons of characteristic frequencies measured from the power spectrum suffer from the relative complexity and potential ambiguity in modelling the power spectrum in terms of multiple Lorentzian-shaped features, where it is not always clear what features should be compared, so that extensive comparison of large samples is required to track the components \citep{KleinWoltvanderKlis2008}.  Furthermore, frequency information alone does not convey the relative contribution of a signal to the variability, which is described by its rms.

\citet{MunozDariasetal2011} developed the `rms-intensity diagram' (RID) which uses the {\it integrated} (0.1--64~Hz) rms as an alternative to the spectral hardness ratio, to study the evolution of rms and source intensity throughout an LMXB outburst.  Given the known correlation between rms and spectral hardness, the RID offers a simple way to track outburst evolution and compare neutron star and black hole sources, which otherwise show spectral differences due to the presence or absence of the boundary layer.  Using the RID approach, \citet{MunozDariasetal2014} showed that the outburst evolution of integrated rms is remarkably similar in neutron stars and black holes.  Subsequently, \citet{Mottaetal2017} used a combination of RID and QPO frequency data to show that the QPOs seen in different parts of the CCD of Z-sources, the flaring, normal and horizontal branch oscillations, can be identified with (respectively) the type A, B and C QPOs in accreting black holes.

The integrated rms is a useful probe of the overall variability amplitude of a source but it does not give any indication of changes in power-spectral {\it shape}, which are observed as the characteristic frequencies {\it and} rms of power-spectral components evolve during an outburst.  It is therefore useful to consider a simple, model-independent approach to showing those changes.  The analogue in spectral studies is the CCD, which reveals broad changes in energy spectral shape.  To this end, \citet{Heiletal2015} developed the `power colour-colour' (PCC) diagram, which quantifies the shape of a power spectrum in terms of only two numbers, by plotting the ratios of integrated power (rms-squared) measured from two pairs of frequency ranges.  \citet{Heiletal2015} showed that BH LMXBs follow a distinct elliptical track in the PCC diagram, corresponding primarily to the transition from hard through hard-intermediate then soft-intermediate state and then to the soft state, which overlaps with the hard state in the PCC diagram due to their similar broad power-spectral shapes, although the integrated rms values of both states are distinct.  \citet{Heiletal2015} further showed that black hole high mass X-ray binary Cyg X-1 follows a similar track in the PCC diagram to the BH LMXBs, and further that the neutron star atoll source Aquila~X-1 shows a similar but slightly offset track to the hard to intermediate part of the black hole PCC diagram.  Thus, given the other timing similarities between the neutron star and black hole systems, it is interesting to compare the PCC behaviour of a much wider sample of accreting neutron star LMXBs, including atoll and Z sources as well as accreting (millisecond) pulsars.

In this paper, we present a PCC analysis of a large and representative sample of accreting neutron star LMXBs which were observed extensively by {\it RXTE}.  We first describe our sample and data analysis approach in Section~\ref{sec:data_analysis}.  In Section~\ref{sec:results} we present our results, first showing the PCC diagram for neutron star LMXBs and a comparison with spectral-evolution using a single-parameter representation of position in the PCC diagram, the `hue' \citep{Heiletal2015inc}.  We compare these diagrams with the behaviour of black hole LMXBs, and then examine their dependence on the Atoll/Z classification of sources and the spin and presence of a magnetic field (identified from a subsample of accreting millisecond X-ray pulsars).  We discuss our results throughout that section and end with our conclusions in Section~\ref{sec:conclusions}.

\section{LMXB sample and data analysis}
\label{sec:data_analysis}

\subsection{Data Selection}
In order to conduct a systematic analysis of accreting black hole and neutron star LMXB variability, a large number of representative observations are needed. To this end, the \textit{RXTE} archival database was used, accessed via the \textit{HEASARC} online service.  Sorting observed accretion-powered neutron star LMXBs by the highest number of observations allowed an initial selection of these systems to be made, before checking it against the literature to ensure the selection covered a wide variety of system types. This allowed for a range of atoll and Z sources to be included, as well as accreting (millisecond) pulsars, which we consider as a separate class and do not further separate into atoll or Z sources (although such distinctions can exist).  Since our aim is to survey accretion-driven variability in neutron stars, we specifically excluded the dipping source EXO~0748-676 from our survey due to its complex absorption-induced variability, despite it having a large number of \textit{RXTE} observations. We also excluded IGR~J17480-2446 from our sample, since its unusual variability is often dominated by mHz flaring thought to be associated with thermonuclear burning rather than accretion variability \citep{Chakraborty&Bhattacharyya2011}.

We also visually inspected all of the {\it RXTE} Proportional Counter Array (PCA) background-subtracted \textsc{standard1} light curves (full $\sim2$--60~keV energy range with 0.125~s time resolution) in our sample for X-ray bursts (distinguished by their clear `fast rise, exponential decay' flux profiles) as well as a few additional cases of dipping (in Aql~X-1, previously reported by \citealt{galloway2016intermittent}) and removed these ObsIDs from our sample.  We note that this is a conservative approach compared to simply excising the times of the bursts, since it also removes the possibility of our results being affected by the bursts physically influencing the accretion-related variability in the surrounding time ranges (e.g. see \citealt{Degenaaretal2018}
for an extensive discussion of the effects of bursts on persistent spectral and variability properties of neutron star LMXBs).

Using the power colour-colour diagram presented in \citet{Heiletal2015} for black hole LMXBs, three representative black hole systems were chosen for comparison, which show good coverage across a full range of accretion states.  For consistency we applied to these systems the same extraction and analysis procedure as used for the neutron stars, rather than reusing the original measurements from \citet{Heiletal2015}.  An overview of the selected objects can be seen in Table~\ref{tab:sources}, including information on the object classification and neutron star spin (if measured).  Note that although \citet{Heiletal2015inc} carried out a study of binary orbit inclination effects on the black hole PCC diagram, we are unable to do so here, due to the much sparser availability of robust binary orbit inclination data from neutron star LMXBs.

\begin{table*}
\caption{Overview of LMXBs showing object classification, neutron star spin frequency (if available) and observation details, separated by compact object and then sorted by name. Systems are divided into atolls (A), Z sources (Z), accreting pulsars (AP), accreting millisecond pulsars (AMP), and objects showing characteristics of both atoll and Z sources (AZ). A further division is made between sources with spin frequencies determined by burst oscillations (B), intermittent pulsations (I) and those determined by persistent accretion-powered pulsations (P) \citep[see][for a review]{watts2012thermonuclear}. Intermittent sources have been assigned to the pulse or burst group on basis of their timing properties \citep{vandoesberghetal2017}. `\#Good' gives the total number of observations without bursts and which show a significant variance detected at a 3$\sigma$-level in all four power colour frequency bands. The total number of available ObsIDs in the \textit{RXTE} archive are also given per source. The final column lists references for the (up to three) filled-in source property columns, which are given at the bottom of this table.}
\label{tab:sources}
\begin{tabular}{cccccccc}
\hline
Source&Type&Burst/Pulse&Spin Freq. (Hz)&\#Good&\#ObsID&References\\
\hline
4U 0614+09 & A & B & 415 & 60 & 502 & 1,2,2 \\
4U 1636-53 & A & B & 581 & 2 & 1556 & 3,4,4 \\
4U 1702-43 & A & B & 329 & 13 & 210 & 5,6,6 \\
4U 1705-44 & A & & & 23 & 516 & 7 \\
4U 1728-34 & A & B & 363 & 10 & 405 & 5,8,8 \\
Aql X-1 & A & I/B & 549 & 123 & 596 & 3,9,10 \\
Cyg X-2 & Z & & & 148 & 567 & 7 \\
GX 17+2 & Z & & & 8 & 206 & 7 \\
GX 340+0 & Z & & & 11 & 97 & 7 \\
GX 349+2 & Z & & & 3 & 142 & 7 \\
GX 5-1 & Z & & & 4 & 167 & 7 \\
HETE J1900.1-2455 & AMP & I/P & 377 & 120 & 361 & 11,9,11 \\
IGR J00291+5934 & AMP & P & 598 & 41 & 479 & 12,13,13 \\
IGR J17498-2921 & AMP & P & 401 & 1 & 129 & 14,9,15 \\
KS 1731-260 & A & B & 524 & 13 & 82 & 5,9,16 \\
SAX J1808.4-3658 & AMP & P & 401 & 17 & 1337 & 17,9,17 \\
SWIFT J1756.9-2508 & AMP & P & 182 & 19 & 50 & 18 \\
Sco X-1 & Z & & & 49 & 598 & 7 \\
Sgr X-1 & A & & & 13 & 109 & 7 \\
Sgr X-2 & Z & & & 51 & 88 & 19 \\
V4634 Sgr & A & & & 68 & 1008 & 20 \\
XB 1254-690 & A & & & 1 & 94 & 21 \\
XTE J0929-314 & AMP & P & 185 & 6 & 46 & 22,22,22 \\
XTE J1701-462 & AZ & & & 96 & 872 & 21 \\
XTE J1751-305 & AMP & P & 435 & 12 & 274 & 23,23,23 \\
XTE J1807-294 & AMP & P & 190 & 2 & 112 & 24,25,25 \\
XTE J1814-338 & AMP & P & 314 & 3 & 93 & 25,9,26 \\
\hline
GX 339-4 & BH & & & 396 & 1401 & 27 \\
H1743-322 & BH & & & 116 & 558 & 28 \\
XTE J1550-564 & BH & & & 158 & 423 & 29 \\
\hline
\end{tabular}

\vspace{1ex}
\raggedright \emph{1}~\citet{mendez1997kilohertz} \emph{2}~\citet{strohmayer2008discovery} \emph{3}~\citet{liu2001catalogue} \emph{4}~\citet{strohmayer1998on} \emph{5}~\citet{galloway2008thermonuclear} \emph{6}~\citet{markwardt1999observation} \emph{7}~\citet{HasingervanderKlis1989} \emph{8}~\citet{strohmayer1997363} \emph{9}~\citet{watts2012thermonuclear} \emph{10}~\citet{zhang1998millisecond} \emph{11}~\citet{watts2009discovery} \emph{12}~\citet{galloway2005discovery} \emph{13}~\citet{markwardt2004orbit} \emph{14}~\citet{papitto2011discovery} \emph{15}~\citet{linares2011rxte} \emph{16}~\citet{smith1997rossi} \emph{17}~\citet{wijnands1998millisecond} \emph{18}~\citet{krimm2007discovery} \emph{19}~\citet{Fridrikssonetal2015} \emph{20}~\citet{van2005relations} \emph{21}~\citet{bhattacharyya2007timing} \emph{22}~\citet{galloway2002discovery} \emph{23}~\citet{markwardt2002discovery} \emph{24}~\citet{markwardt2003discovery} \emph{25}~\citet{markwardt2003xte} \emph{26}~\citet{strohmayer2003xray} \emph{27}~\citet{WijnandsvanderKlis1999} \emph{28}~\citet{homan2005high} \emph{29}~\citet{Homanetal2001}
\end{table*}

\subsection{Data Extraction}
To analyse the large quantity of data, we developed the \textsc{chromos} pipeline\footnote{All software scripts, as well as the results from in this paper, can be found at \url{https://github.com/davidgardenier/chromos}} to link together extraction routines provided in \textsc{ftools} \citep{Blackburn1995}, with our own software for conducting timing analysis.  To obtain signal-to-noise suitable for timing analysis, we only used data from the {\it RXTE} PCA.  We used \textsc{event} mode data where available, otherwise using \textsc{binned} mode data, requiring a time-resolution of $1/128$~s or better and (consistent with the approach of \citealt{Heiletal2015}) an extracted energy range as close to 2-13~keV as possible (determined using the PCA energy-channel conversion table\footnote{\url{https://heasarc.gsfc.nasa.gov/docs/xte/e-c_table.html}}, with channel-ranges selected using the channel binning given in the headers of \textsc{event} or \textsc{binned} mode files).  We also ensured that the lowest energy channels were omitted in \textsc{binned} mode extractions, if the bitsize could have caused overflow errors \citep[see][]{Gleissneretal2004}. This step was necessary for less than ten observations.

Background files were created using the \textsc{ftool} \textsc{pcabackest}.  For sources showing a net count rate larger than 40~ct/s/PCU, the `bright' background model is used, otherwise we use the faint source model.  Good Time Intervals (GTIs) were created using standard pointing criteria, with source elevation above the Earth's limb $>10^\circ$ and pointing offset $<0.02^\circ$. To mitigate any systematic errors in background subtraction, times $<10$~minutes since the last South Atlantic Anomaly (SAA) passage were removed.  To further reduce background contamination for sources with total count rates per PCU $<500$~count~s$^{-1}$, we also selected on electron rate, excluding times with electron rates $>0.1$.  Brighter sources can also lead to higher electron rates even in the absence of high backgorund, hence this criterion is only applied to the lower count-rate sources.  Finally, using information from standard filter files, the times at which a change in number of PCUs occurs were noted, allowing for 32~s either side of these transitions to be filtered during extraction. This prevents any surge, or change in electrical current, from contaminating the count rate.

Light curves or spectra can subsequently be extracted using the \textsc{ftools} \textsc{saextrct} and \textsc{seextrct}.  The extracted light curves combine data from all available Proportional Counter Units (PCUs).  The time resolution for light curves is set to be $1/128\ $s. Background light curves with 16~s resolution are also extracted from the background files using identical time-selections as the high time-resolution light curves. Spectra (for hardness-ratio determination, see below) are extracted from \textsc{standard2} PCA files and corresponding background files, again using the same time-selections as the high time-resolution light curves. For consistency and because it has a reasonably stable response, we only use PCU 2 for spectral extraction. \\

\subsubsection{Timing Analysis}
Light curves are background corrected, interpolating between consecutive background data points to obtain $1/128\ $s resolution. Power spectra are computed using discrete Fourier transforms \citep[see][]{Uttleyetal2014}. Following the procedure given in \citet{Heiletal2015}, we take discrete Fourier transforms of continuous 512~s segments of an observation before averaging and normalising to units of fractional variance per Hz \citep{BelloniHasinger1990}. Associated errors on the power spectrum are calculated by dividing each power by $\sqrt{M}\hspace{2pt}$, with $M$ the total number of segments in the observation. The unbinned powers of any noise process are drawn from a scaled $\chi^2_2$ distribution, but errors on the power spectrum can be approximated as Gaussian provided that a large number of samples are binned.  Power spectra are subsequently corrected by subtracting the constant Poisson noise level (e.g. \citealt{Uttleyetal2014}), applying a slight rescaling to the noise level to account for dead-time \citep{Jahodaetal2006}.
\begin{figure*}
 \includegraphics[width=\textwidth]{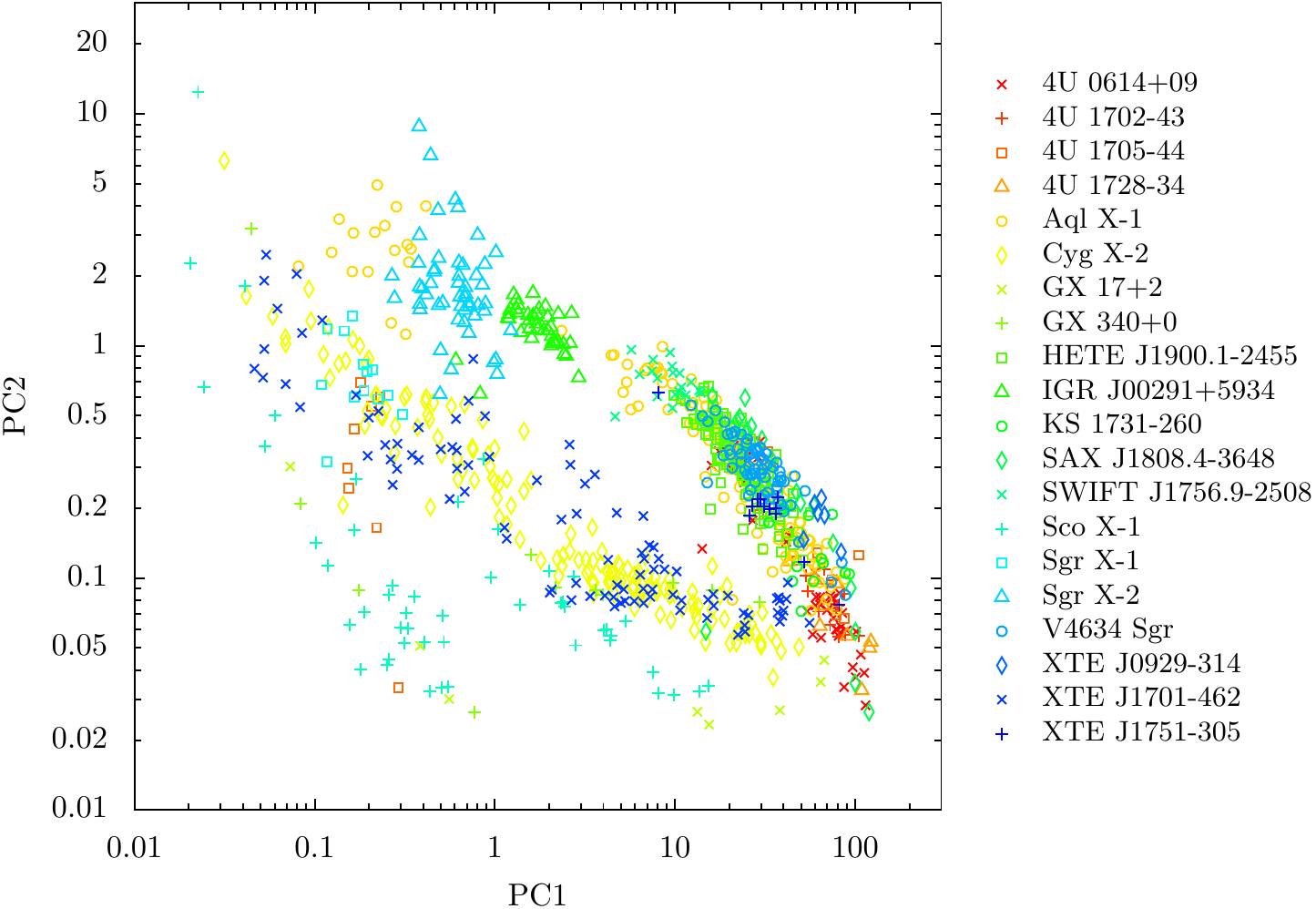}
 \caption{A power colour-colour (PCC) diagram showing tracks for neutron star LMXBs, with PC1 defined as the variance ratio for (0.25-2.0~Hz)/(0.0039-0.031~Hz) and PC2 as the variance ratio for (0.031-0.25~Hz)/(2.0-16.0~Hz). For clarity, errors are not shown, but are on average $\sim$17\% of the PC value.  While providing an overview of the general trend, tracks of individual objects can be best followed in appendix~\ref{sec:pcc}, where individual PCC~diagrams are shown for each system (with error bars included).}
 \label{fig:pc_all_ns}
\end{figure*}

Following \citet{Heiletal2015} we determine power colours for each observation from their averaged power spectra. First, variances can be calculated by integrating the power over four frequency bands with the same (factor 8) geometric spacing; A: 0.0039-0.031~Hz, B: 0.031-0.25~Hz, C: 0.25-2.0~Hz and D: 2.0-16.0~Hz.   Denoting the measured variance as $V_X$ (where $X$ is the given band), we define two power colours as:
\begin{equation}
 PC1 = \frac{V_C}{V_A} \hspace{20pt} \textrm{and} \hspace{20pt} PC2 = \frac{V_B}{V_D}
\end{equation}
These measurements allow an observation to be placed in a PCC~diagram, with PC1 on the horizontal axis and PC2 on the vertical axis. Errors are propagated from those on the variance, calculated as described in \citet{Heiletal2012}.  Only PCC points with a positive variance detected at $>3\sigma$ in all frequency bands are used in subsequent analysis, to ensure an accurate positioning in the PCC~diagram.\\

\subsubsection{Spectral hardness determination}
\label{sec:hardness}
We also use the {\it RXTE} PCA spectrum for each observation to calculate a corresponding spectral hardness, for comparison with the timing evolution measured using the power colours.  The \textsc{ftool} \textsc{pcarsp} is run for all observations to determine an instrument response matrix for each observation (taking into account evolution of the response through the lifetime of the mission).  We then determine the hardness ratio in as model- and instrument-independent a way as possible by unfolding the energy spectrum around a constant using the \textsc{xspec} software \citep{Arnaud1996} (equivalent to dividing by the instrument effective area).  The resulting energy spectrum is used to calculate the energy spectral hardness by taking the ratio of 9.7-16.0~keV integrated (energy) flux to the 6.4-9.7~keV integrated flux (interpolating where spectral channels do not exactly match the chosen energy ranges).  These energy ranges were chosen in order to compare results with the bands used to define the `hard colour' in previous neutron star studies \citep[e.g.][]{Gladstoneetal2007}.  Furthermore, this choice of harder energy ranges avoids a significant disk blackbody contribution to the spectrum, which would complicate interpretation of any spectral changes when considered together with the boundary layer and power-law components.  Several tests were conducted on the effects of differing hardness ratio energy bands on, for instance, the hardness-hue~diagram. While varying the defined energy bands did have the effect of stretching the range of hardness values, no benefit was found in changing the hardness ratio energy bands from the values used in prior studies.

\begin{figure}
 \includegraphics[width=\columnwidth]{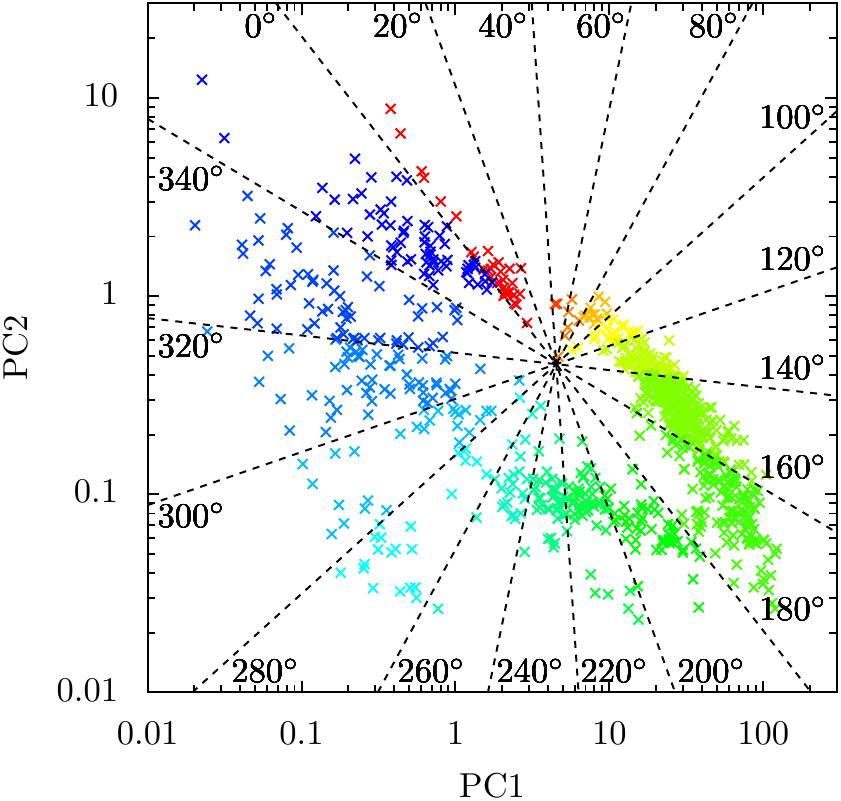}
 \caption{A PCC~diagram showing the division of $20^\circ$ hue bins for neutron star LMXBs. The starting angle is defined as the angle at $45^\circ$ in a counter-clockwise direction from the vertical axis.  The colours are chosen to map on to the different bins, the same scheme is also used to illustrate the power spectra in Fig.~\ref{fig:ps_overview}.  A similar figure for black holes can be seen in Fig.~2b from \citet{Heiletal2015}.}
 \label{fig:pc_hue_bins}
\end{figure}

\section{Results and discussion}
\label{sec:results}
\subsection{The power colour-colour diagram and power-spectral hue of neutron stars}
\label{sec:pcc_hh_ns}
Applying \chromos to the population of neutron stars given in Table~\ref{tab:sources} reveals that most objects follow similar tracks in the power colour-colour (PCC) diagram, and that the morphology of these tracks is very similar to that seen by \citet{Heiletal2015} for black holes. This trend can be seen in Fig.~\ref{fig:pc_all_ns}, where a distinct elliptical shape emerges from the plotted data. In this plot, following the approach of \citet{Heiletal2015}, PC1 is defined as the variance ratio (0.25-2.0~Hz)/(0.0039-0.031~Hz) and PC2 as the variance ratio (0.031-0.25~Hz)/(2.0-16.0~Hz). Only observations with a significant ($>3\sigma$-level) detected variance in all four power colour frequency bands have been included in the diagram, with the additional exclusion of objects with 5 or fewer PCC points. For the sake of clarity, error bars have been omitted, typically being around 17\% of the given power colour values. PCC tracks for individual objects (including error bars) can be found in appendix~\ref{sec:pcc}, in which each object has been plotted for comparison with all other neutron star systems.

\begin{figure*}
 \includegraphics[width=\textwidth]{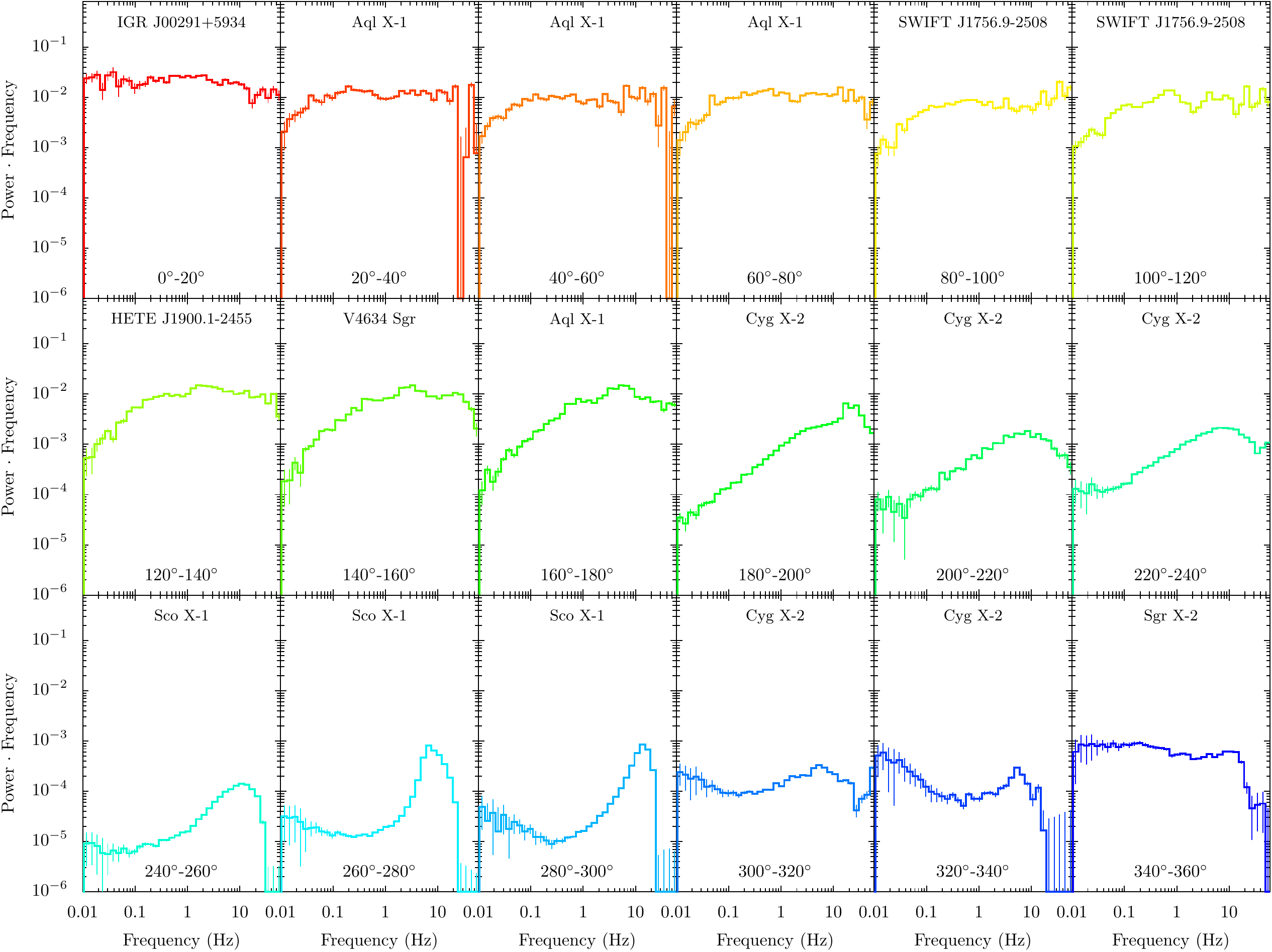}
 \caption{An overview with representative examples of neutron star power spectra per hue bin, as seen in the PCC~diagram in Fig. \ref{fig:pc_hue_bins}. Each graph shows a representative power spectrum from the most common object within the hue bin in order to trace power spectral evolution through various states.  The colours for each plotted power spectrum are chosen to map on to the colours shown for each hue bin in Fig.~\ref{fig:pc_hue_bins}.  Power-spectra have been rebinned to equal-sized bins in log-frequency to reduce the scatter at high frequencies.  Appendix~\ref{sec:ps} contains additional information, showing a larger selection of power spectra per hue bin.}
 \label{fig:ps_overview}
\end{figure*}

While power colours are useful for comparing the evolutionary tracks of system timing properties, they require two dimensions (PC1 and PC2) to classify a system in terms of its overall power-spectral shape. Reducing this scheme down to a single parameter can be helpful in comparing the evolving power-spectral shape of a system against other parameters. To this end, the `hue'~parameter can be introduced \citep{Heiletal2015,Heiletal2015inc}. Defined as the angle of a point in the PCC~diagram with respect to a central point, hue runs from $0^\circ$ to $360^\circ$ in a clockwise direction, starting from a line in the northwest direction. Following the original defining locus given in \citet{Heiletal2015}, a central point with the coordinates (4.51920, 0.453724) is chosen as reference point. Dividing neutron star PCC tracks into hue bins of $20^\circ$, as seen in Fig.~\ref{fig:pc_hue_bins}, allows an overview of representative examples from each hue bin to be created, showing the power spectral evolution throughout the PCC~diagram, which is shown in Fig.~\ref{fig:ps_overview}.  Appendix~\ref{sec:ps} increases this sample, by comparing neutron star power spectra with black hole power spectra at various hues. While keeping GX~339-4 power spectra in the top left of each panel, going from left to right, top to bottom, these panels show power spectra from the most common objects within each hue bin.

As expected given the similar shapes in the PCC diagram, the power-spectral evolution of neutron star LMXBs through the diagram is similar to what is seen in the black hole systems (see Fig.~2c of \citealt{Heiletal2015} and accompanying discussion in their Section~3.1).  The evolution from top-left to bottom-right sides of the elliptical track (along a well-defined {\it `upper track'} on the top-right side of the ellipse) corresponding to hue bins from 0--200$^{\circ}$, is due mainly to power being removed from the lower-frequencies so that the overall shape becomes more `band-limited' in the higher frequency range.  Then, for hue bins from 200$^{\circ}$ to larger values (where the `track' becomes much more scattered and diffuse, forming a {\it `lower cloud'} of points) there remains a significant peak in high-frequency power but power-law like low-frequency noise appears, which pushes the overall shape closer to the very broad noise component observed for low hue values.  It is interesting to note that an overall sharp drop in power-spectral normalisation accompanies the transition from upper track to lower cloud - this would not affect the power-colours but would produce significant evolution in the RID, and is thus clearly linked to the transition to the softer states \citep{MunozDariasetal2014}.

\begin{figure*}
 \includegraphics[width=\textwidth]{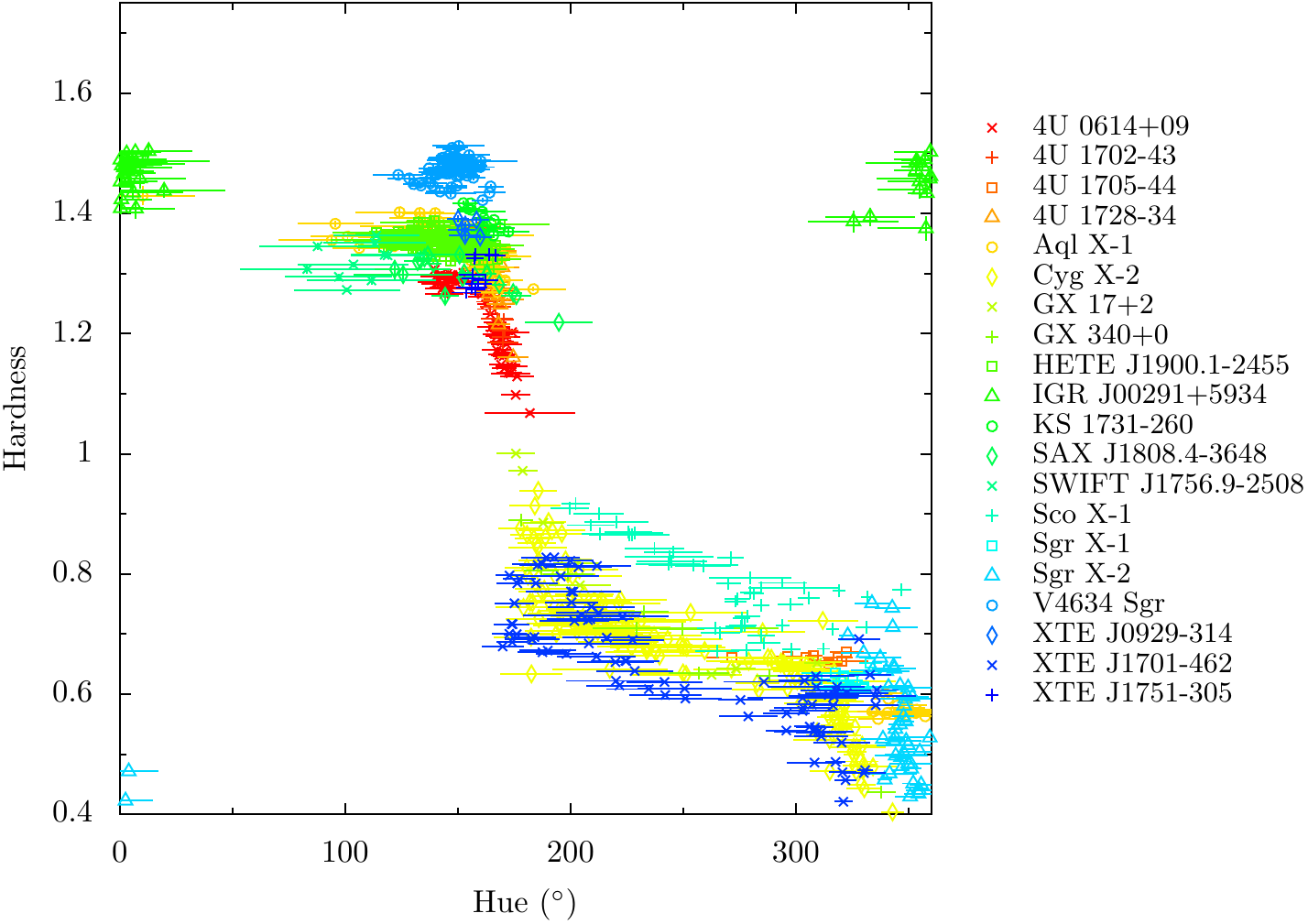}
 \caption{A hardness-hue (HH) diagram showing the evolution of PCC-tracks for neutron star LMXBs through use of the hue.  The hardness is defined as the ratio of 9.7-16.0~keV to 6.4-9.7~keV flux.  See text in Section~\ref{sec:pcc_hh_ns} for an explanation of the definition of hue.  Supplementary HH~diagrams can be found in appendix~\ref{sec:hh}, showing the tracks of individual objects with the tracks of the entire neutron star sample as reference.}
 \label{fig:hh_all_ns}
\end{figure*}

The overall similarity in neutron star and black hole LMXB PCC diagram shapes, suggests that a similar mapping of broadband power-spectral shape to spectral state may be made in both kinds of system (see Fig.~2a in \citealt{Heiletal2015}).  The upper track corresponds to the evolution from the hard state to the hard-intermediate state, while the lower cloud corresponds to the change to soft-intermediate through to the soft state.  As is the case for black holes, the soft and hard states overlap significantly in the top left corner of the diagram, since they have similar broad power-spectral shapes.  We can check the correspondence to spectral state by comparing with the evolution of the energy spectrum.  Comparing hue with energy spectral hardness provides a simple way to compare changes in power-spectral shape with changes in the observed energy spectrum. A hardness-hue (HH) diagram for neutron star LMXBs can be seen in Fig.~\ref{fig:hh_all_ns}, with (as described in Section~\ref{sec:hardness}) the hardness defined as the ratio of the flux in 9.7-16.0~keV to the flux in 6.4-9.7~keV.   Next to the selection methods for PCC~points described in the first paragraph of this section, only points with a hue-error $<\!30^\circ$ are included in this diagram, where errors are propagated through from the PCC~errors. In a similar fashion to the PCC~diagrams in appendix~\ref{sec:pcc}, individual HH~diagrams can be found in appendix~\ref{sec:hh}, allowing the evolution of an object within a HH~diagram to be traced against that of the other neutron star LMXBs.

The neutron star LMXB HH~diagram shows clearly (and similarly to that reported for black holes by \citealt{Heiletal2015inc}) that low hue values correspond to hard spectra and that from 150--200$^{\circ}$ (corresponding to the hard to soft-intermediate state transition inferred from the PCC diagram) there is the expected clear change in energy spectral shape. Thus the spectral state transition for neutron stars is also connected to hue in the same way as for black holes.

\subsection{Systematic differences between neutron star and black hole LMXBs}
\begin{figure*}
 \centering
 \begin{subfigure}{0.51\textwidth}
  \includegraphics[width=\textwidth]{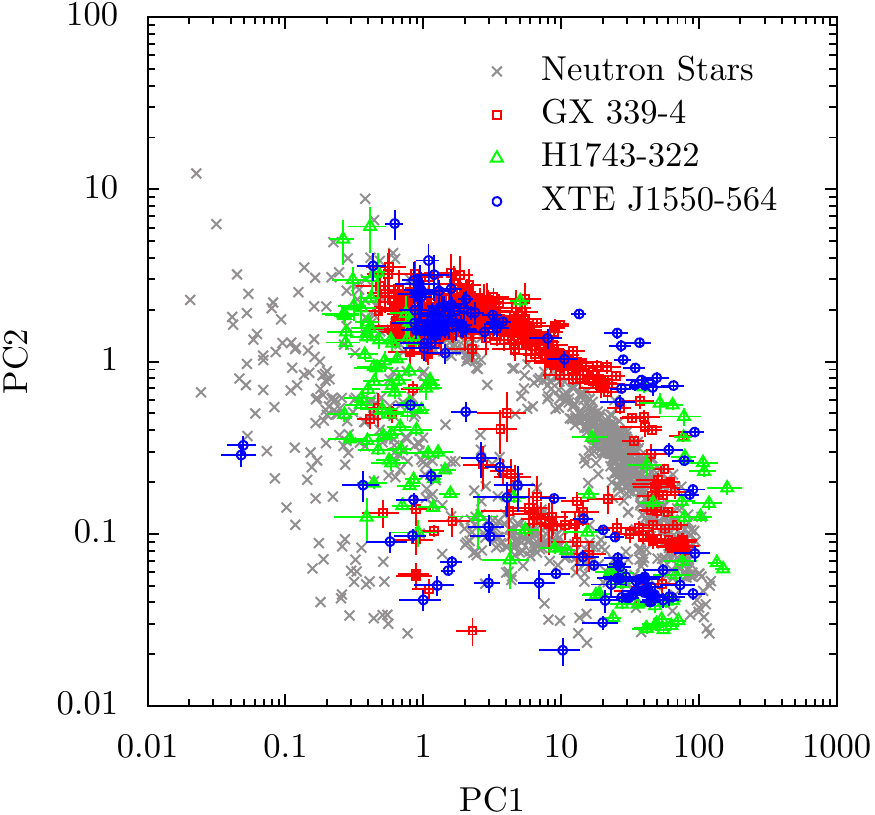}
 \end{subfigure}%
 \begin{subfigure}{0.49\textwidth}
  \includegraphics[width=\textwidth]{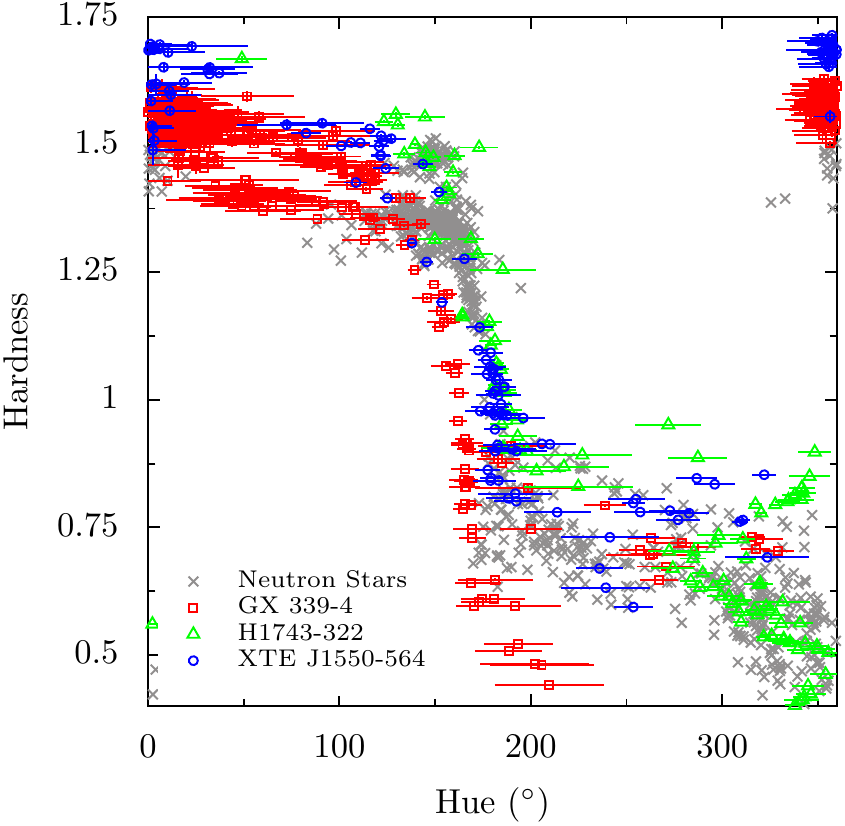}
 \end{subfigure}
 \caption{\emph{left} A PCC~diagram showing black hole systems, with in grey crosses the neutron star PCCs as given in Fig.~\ref{fig:pc_all_ns} as reference. \emph{right} The same systems plotted in a HH~diagram.}
 \label{fig:ns_bh}
\end{figure*}

Having established the general trend of neutron star PCC tracks and hue, and the qualitative comparison with the results on black holes from \citet{Heiletal2015,Heiletal2015inc}, it follows to compare these tracks directly with those of black hole systems and look for differences.  In the left panel of Fig.~\ref{fig:ns_bh}, three representative transient black holes have been plotted for comparison with neutron stars, with additional information on these systems given in Table~\ref{tab:sources}. Both types of system show similar paths, as expected from our qualitative comparison in the previous section, yet a clear distinction is found for the upper track, which for the black holes is shifted to the right with respect to the neutron stars. In the right panel of Fig.~\ref{fig:ns_bh}, a HH~diagram is shown for the same systems, where the hardness is classified as the same flux ratio of (9.7-16.0~keV)/(6.4-9.7~keV). With the hue washing out any radial differences in PCC position, in the HH~diagram the black holes closely follow the neutron stars albeit with better coverage of low-hue angles (which are mostly removed for neutron stars due to the lower signal-to-noise in their hard state) and notably more spread in which hue angles correspond to the spectral state change. \\

\begin{figure*}
 \includegraphics[width=\textwidth]{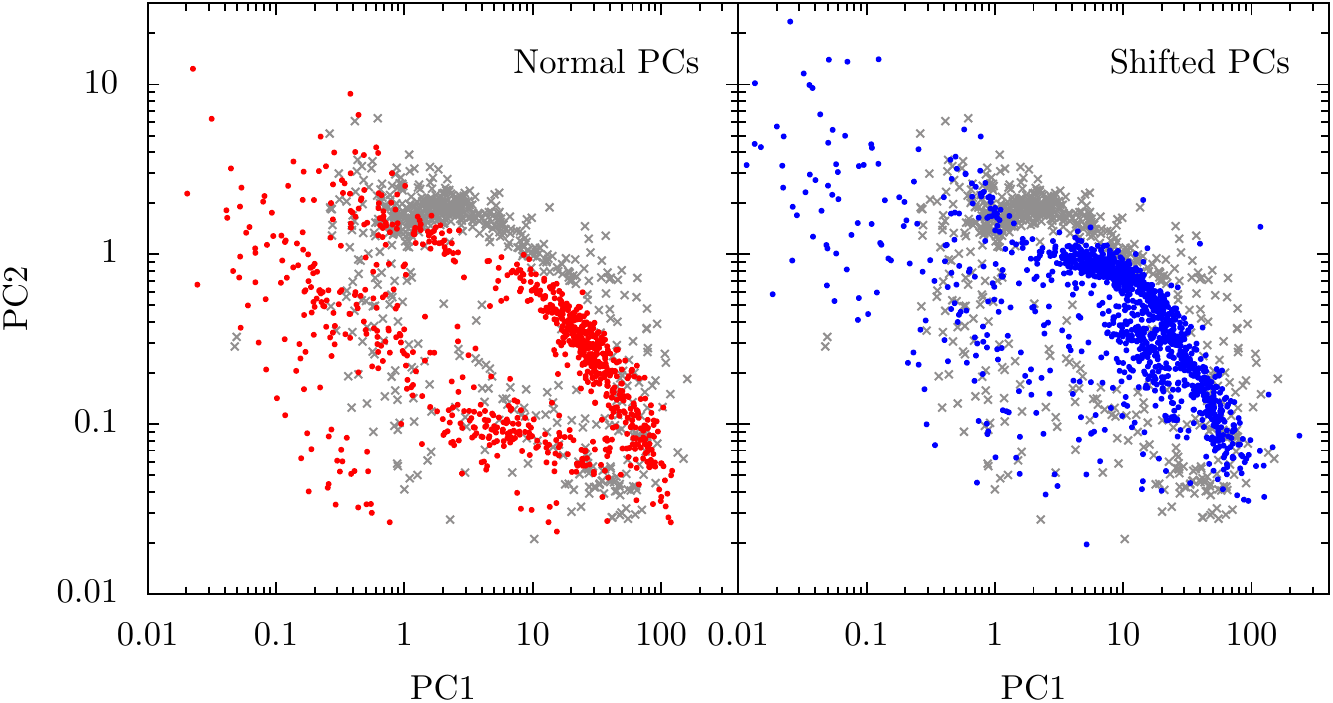}
 \caption{\emph{left} A PCC~diagram showing neutron stars in red dots against black hole systems in grey crosses. \emph{right} PCCs for neutron stars where the frequency bands have been shifted up by a factor of four, given in blue dots. The black hole systems in grey crosses have retained the original frequency bands for their PCC values.}
 \label{fig:shiftedpc}
\end{figure*}

We can consider two possible reasons for the differences between black hole and neutron star PCC tracks: mass-scaling of the power spectrum, or systematic differences in power-spectral shape.  Firstly, while black hole and neutron star LMXBs have similar broad-band power-spectral shapes, the frequency at which power-spectral features occur can be a factor five lower for black holes than for neutron stars \citep{kleinwolt}, with lower mass black hole systems showing a slightly smaller shift.  The shift in frequencies is probably caused by the systematically lower compact object mass expected for neutron stars, leading to a corresponding rescaling of characteristic time-scales in the innermost regions.  This effect can be approximately accounted for by shifting the frequency ranges for neutron star power colours up by a factor of four in the power spectrum.  This shift changes the contiguous frequency boundaries for integration of the power spectrum to be: 0.0156, 0.125, 1, 8 and 64~Hz. Fig.~\ref{fig:shiftedpc} shows the result of shifting the frequency bands for neutron stars. In the left panel, the original PCC~values for neutron stars can be seen as red dots against the black hole PCC~values as grey crosses. In the right panel are the shifted neutron star PCC~values (blue dots), shown against the unaffected black hole PCC~values (grey crosses). While the shifted PCC~values show a greater overlap between neutron star and black hole PCCs, both tracks can still be distinguished, so the mass-scaling of the power spectrum may not be the only possible explanation for the difference between neutron star and black hole systems.

An additional cause of differences in the neutron star and black hole PCC diagrams may be systematic differences in power-spectral shape that are independent of mass-scaling effects.  For example, \citet{SunyaevRevnivtsev2000} noted that neutron stars show systematically more high-frequency ($\gtrsim 10$--$50$~Hz) power than black holes even {\it after} correcting for mass-scaling of the power-spectral frequencies (see also \citealt{kleinwolt,KleinWoltvanderKlis2008}).  The effect would be to enhance the variance in the highest frequency band and hence suppress the value of PC2 compared to the black hole values, which could explain some of the downward shift of neutron stars compared to black holes seen in the upper track of the PCC diagrams.  A further significant difference could be due to the presence of low-frequency QPOs, which generally appear to be stronger in the black hole systems compared to neutron stars \citep{KleinWoltvanderKlis2008}.  Since these QPOs become more prominent in the intermediate states, their enhanced strength in the black hole systems could further help to explain the deviation between black hole and neutron star systems along the upper track of the diagram, since the effect of stronger low-frequency QPOs is to push the lower part of the track up and to the right (see \citealt{Heiletal2015inc} Fig.~4, which shows the effect on the black hole PCC diagram of removing the QPO contribution to variance).

\citet{Heiletal2015inc} also showed that the location of the black hole PCC diagram upper track appears to depend on binary orbit inclination, with higher inclination (i.e. more edge-on) systems, such as H1743-322 and XTE~J1550-564, showing tracks shifted up and to the right compared to those of lower-inclination systems such as GX~339-4.  \citet{Heiletal2015inc} showed that this inclination dependence disappears once the QPO contribution is removed from the calculation of power-colours, implying that QPO rms amplitude depends on inclination (see also \citealt{Mottaetal2015}) while the shape of the broadband noise does not.  Unfortunately, due to the absence of binary orbit inclination data for most of the systems in our sample, it is not possible for us to do the same comparison with neutron stars.  However, we note for completeness that this inclination-dependence of QPO strength may may systematically enhance the spread in black hole PCC diagram tracks compared to those for neutron stars, due to the relative weakness of the neutron star QPOs.

\subsection{Atolls and Z sources}
\begin{figure*}
 \centering
 \begin{subfigure}{0.51\textwidth}
  \includegraphics[width=\textwidth]{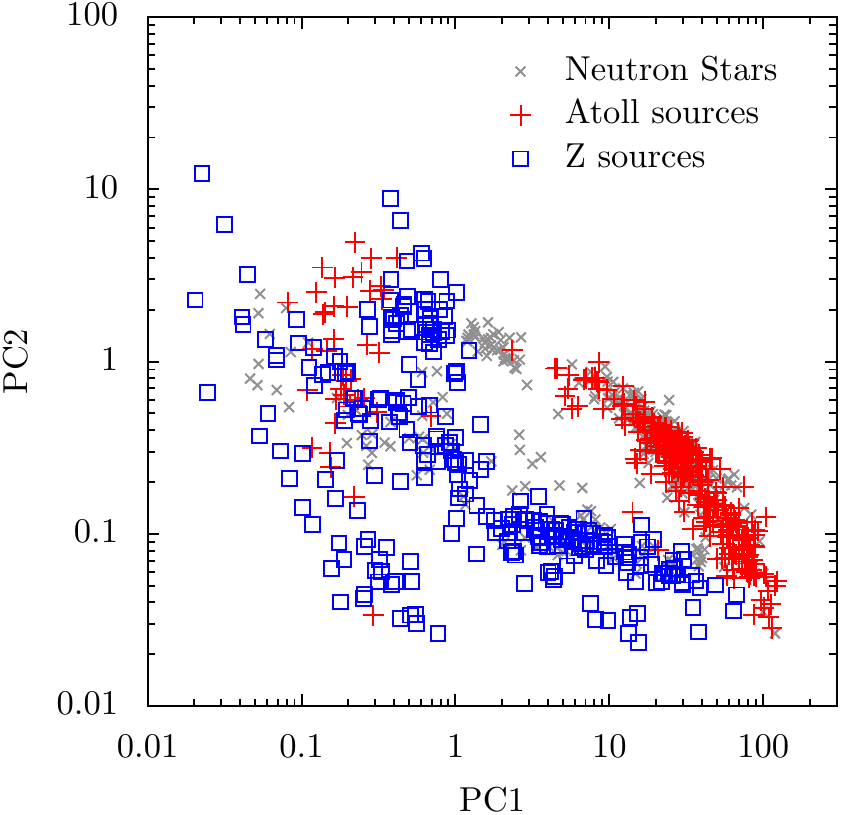}
 \end{subfigure}%
 \begin{subfigure}{0.49\textwidth}
  \includegraphics[width=\textwidth]{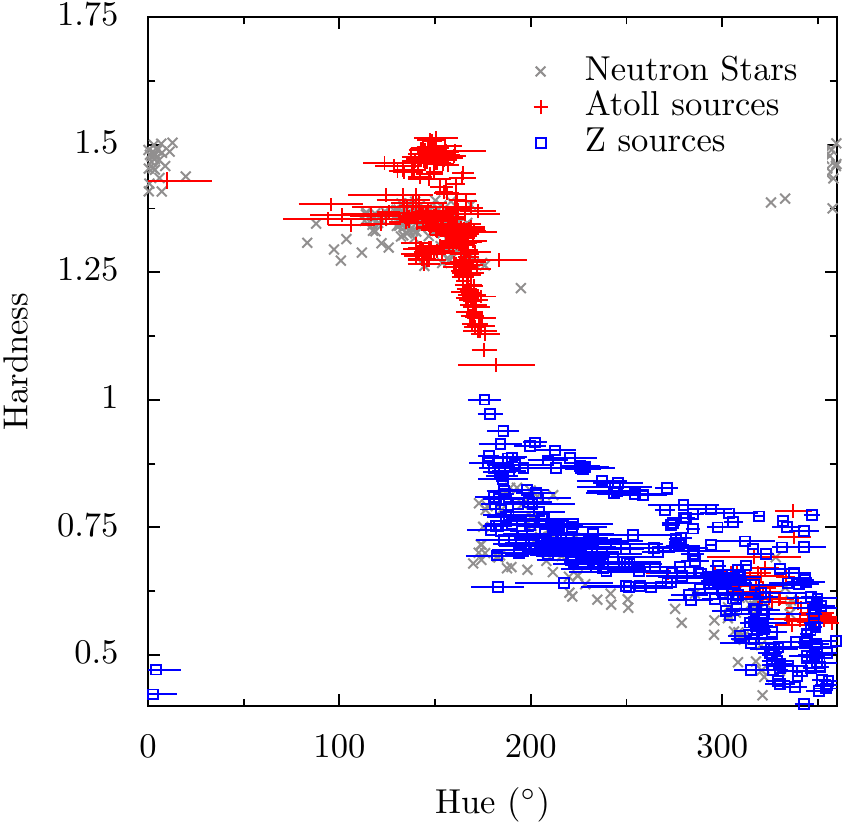}
 \end{subfigure}
 \caption{\emph{left} A PCC~diagram with systems classified as atoll sources in red plusses, Z sources in blue squares and unclassified sources in grey crosses. An overview showing the type of each individual system can be found in Table~\ref{tab:sources}. \emph{right} Replotting the same systems in a HH~diagram.}
 \label{fig:atoll_z}
\end{figure*}

In Fig.~\ref{fig:atoll_z} we show the PCC and HH diagrams for neutron stars with the atoll and Z sources distinguished.  The upper track is populated exclusively by atoll sources, while the lower cloud (and top-left corner of the diagram) is populated by both atoll and Z sources.  The distinction between atoll and Z sources is based on the tracks they follow in the CCD \citep{HasingervanderKlis1989}, as opposed to the hard, intermediate and soft states in black holes, which are identified on the basis of relative spectral hardness and position in the HID.  A number of lines of evidence suggest that the main distinction between atoll and Z sources is one of accretion rate, rather than other neutron star parameters such as the magnetic field strength (e.g. \citealt{Fridrikssonetal2015}).  For example, XTE~J1701-462 (which we list as an unclassified source here) was the first outbursting neutron star LMXB that was seen to transition from being Z source through to an atoll source as its luminosity dropped by an order of magnitude \citep{Linetal2009,Homanetal2010}.

The PCC diagram behaviour is consistent with previous interpretations of the atoll and Z source classes in comparison with black hole systems (e.g. \citealt{Mottaetal2017}), namely that the Z sources correspond to sources in the most luminous (close to the Eddington limit) soft and soft-intermediate states, while atoll sources correspond to hard, intermediate and soft states seen at lower luminosities.  In other words, high luminosity hard states do not seem to exist in neutron star systems.  Crucially, this distinction can now be made using the power-spectral shape alone, i.e. we do not see Z sources with hard-to-hard-intermediate-state-like timing properties (i.e. on the upper track of the PCC diagram).  Therefore, the absence of luminous hard and hard-intermediate spectral states (i.e. at Z-source luminosities) cannot simply be argued to result from the presence of the boundary layer, e.g. with additional seed photons cooling the Comptonising region and softening the spectrum. This distinction is also apparent from the HH diagram, which shows atoll sources with hard spectra and low hue values (corresponding to the upper track), as well as soft spectra and large hue values, while Z sources are only soft and with large hue values.

Finally, we note that Z-sources can further be split according to the detailed shapes of their CCD tracks into Cyg (X-2)-like and Sco (X-1)-like subtypes (also thought to link to accretion rate, with Cyg-like corresponding to higher luminosities).  Due to the limited statistics for comparing just a few sources, we do not explicitly compare the PCC and HCC diagrams of these classes, but the interested reader can use the individual diagrams shown in the Appendix to compare objects.

\subsection{Effects of neutron star spin and magnetic field}
\begin{figure*}
 \includegraphics[width=\textwidth]{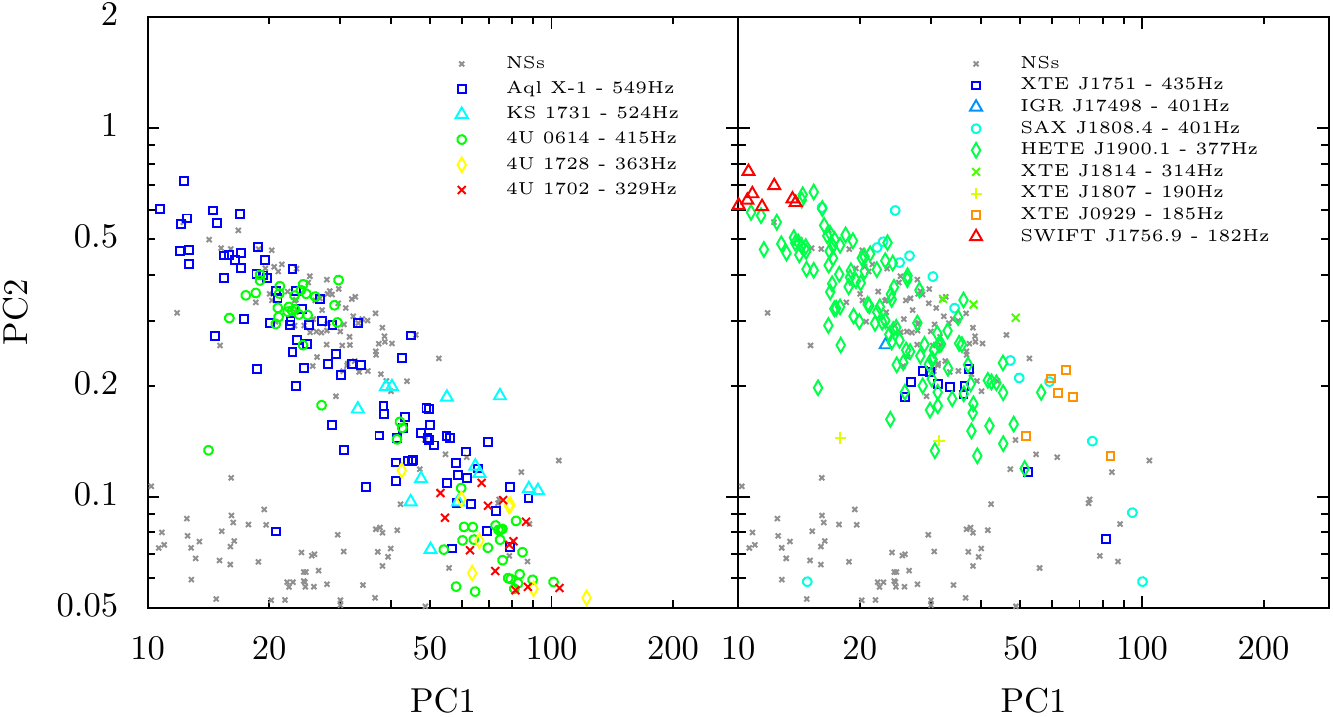}
 \caption{\emph{left} Burst oscillation sources plotted in order of frequency, with neutron star LMXBs without a defined spin frequency given in grey crosses. \emph{right} PCC~diagram showing persistent pulsators together with the PCC values of the other neutron stars.}
 \label{fig:spin}
\end{figure*}

It is interesting to see whether the behaviour on the PCC diagram is affected by the spin of the neutron star, and further whether there is then a difference between the non/weakly-magnetic systems (where spin is established from thermonuclear burst oscillations, e.g. \citealt{chakrabarty2003nuclear,watts2012thermonuclear}) and the magnetic systems, where spin is established from millisecond persistent X-ray pulsations (these sources are the accreting millisecond X-ray pulsars, AMXPs).  While the observed spin frequencies (hundreds of Hz) typically fall far from the frequency bands that power colours probe, some effect might be seen if, for example, the spin affects the accretion flow via the pulsar magnetosphere. Fig.~\ref{fig:spin} shows the well-defined part of the upper track which is well-populated by both AMXPs and bursters, with objects colour-coded according to their spin frequency.  For both types of system we see no clear dependence of the upper track on the spin of the object, nor do we see a clear distinction between the shape of the track and whether the system shows persistent pulsations or only burst oscillations.  Therefore we conclude that the hard-intermediate state power-spectral shape below 16~Hz and its evolution, are not strongly affected by the spin of the neutron star or whether the neutron star is relatively more strongly magnetic.  However, we note for completeness that the inclusion in our sample of accreting ms pulsars means that the magnetic fields we consider are still relatively modest ($\sim 10^{8}$~G, e.g. \citealt{Mukerjeeetal2015}).  The strongly magnetic accreting systems ($\sim 10^{12}$~G) correspond to the more slowly spinning X-ray pulsars seen in HMXBs \citep{Caballero&Wilms2012}, with very different properties to the accretion-powered LMXBs considered here.

\section{Conclusions}
\label{sec:conclusions}
We have carried out a comprehensive model-independent analysis of the evolution of neutron star LMXB timing properties.  The evolution of broadband power-spectral shape, as quantified in a simple diagram which plots a pair of `power colours' - ratios of power integrated over different frequency ranges -  confirms the remarkably strong similarities between neutron star and black hole LMXB timing evolution during outburst and through spectral state transitions.  The neutron star power-colour-colour (PCC) diagram shows similar features to the black hole equivalent reported by \citet{Heiletal2015}, including a well-defined `upper track' which corresponds to the hard through hard-intermediate states, and a less well-defined `lower cloud' corresponding to the hard-intermediate to soft-intermediate to soft state transition.  Among our additional findings are:
\begin{enumerate}
\item Using a single angle, the `hue' to quantify position around the PCC diagram, we construct the `hardness-hue' (HH) diagram and confirm that neutron stars show a clear spectral transition from hard to soft across a narrow range of hues (150$^{\circ}$--200$^{\circ}$) which corresponds remarkably well to the equivalent transition in black holes.
\item The black hole and neutron star PCC diagrams show systematic offsets in the upper track, which are reduced when shifting upwards (by a factor 4) the frequency ranges used to define the neutron star power-colours, as would be expected if characteristic power-spectral frequencies scale inversely with compact object mass.  The remaining differences may be linked to the presence of low-frequency QPOs, which seem to be systematically stronger in the black hole systems, together with the relatively larger amplitude of high-frequency power seen in neutron star systems, which may be linked to the presence of the boundary layer emission, acting to amplify high-frequency signals from the innermost accretion flow.
\item Atoll sources occupy all parts of the PCC diagram, but Z sources only occupy the lower cloud, i.e. Z sources do not appear in hard or hard-intermediate states.  These results are broadly consistent with the identification of the Z-sources with luminous, high-accretion rate soft states (e.g. \citealt{Homanetal2010,Mottaetal2017}), where the absence of luminous states with harder spectral shapes (which are seen for black hole LMXBs) can be attributed to the presence of the boundary layer.  However, it is not clear as to why the equivalent hard and hard-intermediate state timing properties should also be absent, if these are only driven by the structure of the variable accretion flow and not the type of central emission region.
\item Finally we note that, for the accretion-powered LMXBs considered here, there is no obvious systematic dependence on neutron star spin or magnetic field, of the position of the upper track of the PCC diagram (for which a comparison can be reliably made).  This result is expected if the effects of spin and the magnetosphere are confined to the highest-frequency variability associated with the very innermost parts of the accretion flow, which would not impact the power-colours measured in the selected, lower, frequency ranges.
\end{enumerate}
The close similarities in timing evolution of accreting black hole and neutron star systems support the now common idea that timing properties  tell us about the physics of accretion which, after accounting for mass-scaling of characteristic time-scales and the filtering/amplifying effects of the central emission region, is relatively insensitive to the type of compact object.  Thus, following the approach of \citet{Heiletal2015inc} for black holes, the power colours of neutron stars may be used as a proxy for the accretion state and accretion flow structure, largely independent of the spectrum.  This approach can allow more detailed comparative studies of different systems where the effects on the spectrum of other system parameters (such as neutron star spin and binary orbit inclination) can be determined, with degeneracies due to the large-scale state evolution removed using the power colour diagnostic.

\section*{Acknowledgements}
We would like to thank Lucy Heil for useful discussions and for providing comparison data to check our power-colour estimates, and Rudy Wijnands and Jeroen Homan for useful discussions.  DWG acknowledges funding from the European Research Council under the European Union's Seventh Framework Programme (FP/2007-2013) / ERC Grant Agreement n.~617199, and support from the Netherlands Institute for Radio Astronomy (ASTRON) and the Netherlands Foundation for Scientific Research (NWO). This research has made use of data obtained through the High Energy Astrophysics Science Archive Research Center Online Service, provided by the NASA/Goddard Space Flight Center. \textsc{chromos} makes use of Astropy v1.0.3 \citep{astropy}, Ftools \citep{Blackburn1995}, Numpy, Pyx and Scipy.




\appendix

\section{Power Spectra}
\label{sec:ps}

\begin{figure*}
 \centering
 \begin{subfigure}{0.47\textwidth}
  \includegraphics[width=\textwidth]{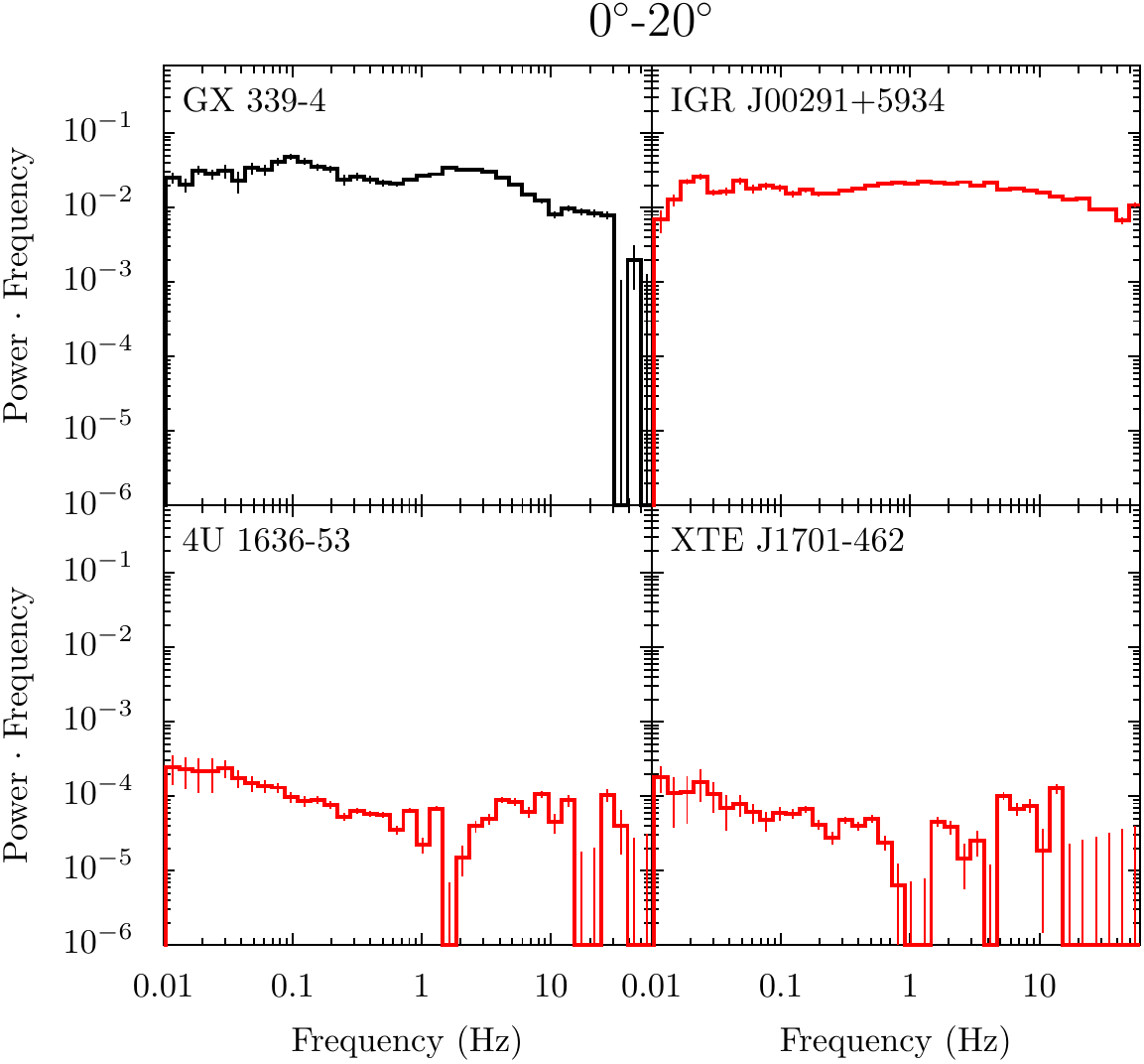}
 \end{subfigure}%
 \begin{subfigure}{0.47\textwidth}
  \includegraphics[width=\textwidth]{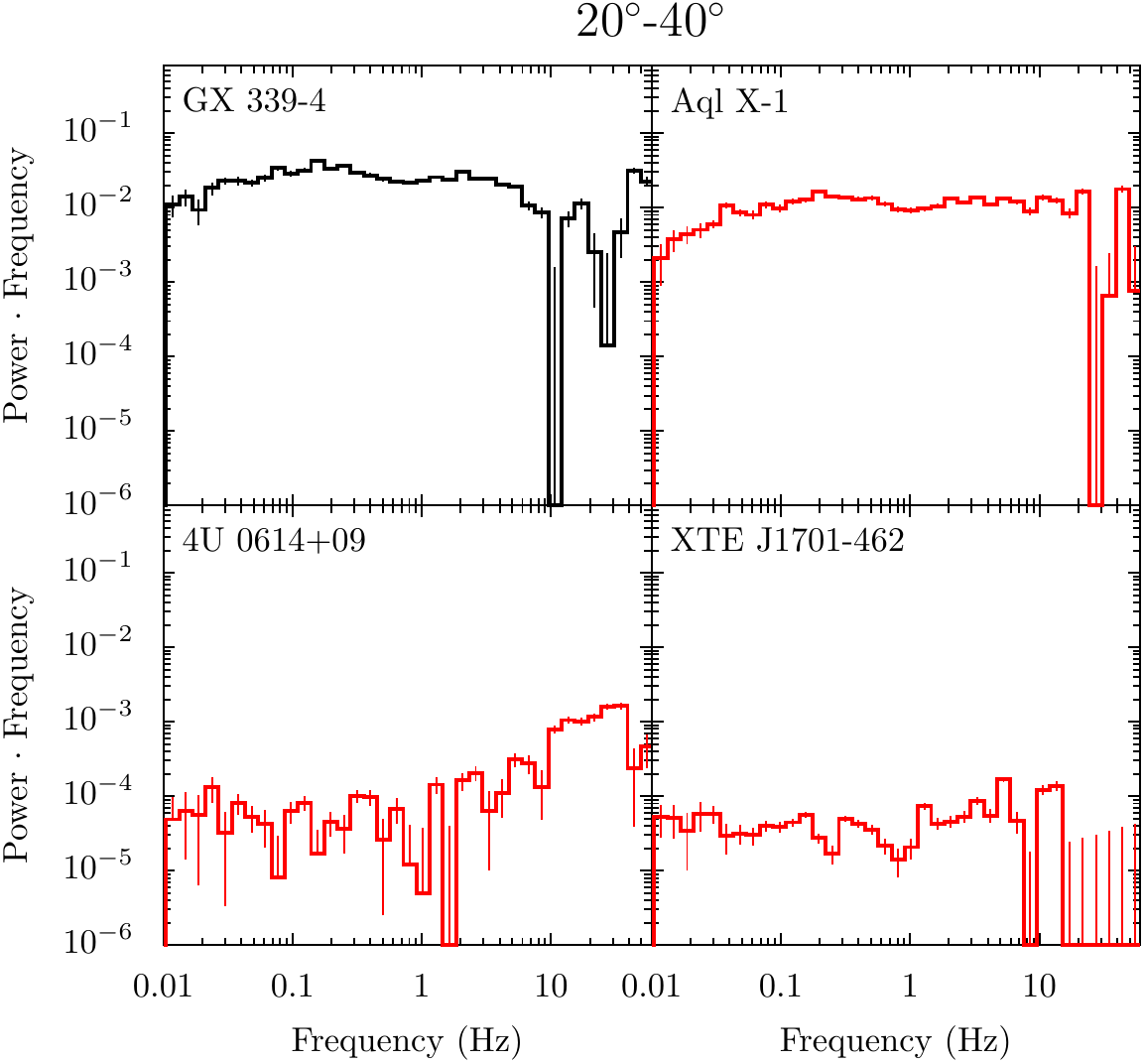}
 \end{subfigure}
 \begin{subfigure}{0.47\textwidth}
  \includegraphics[width=\textwidth]{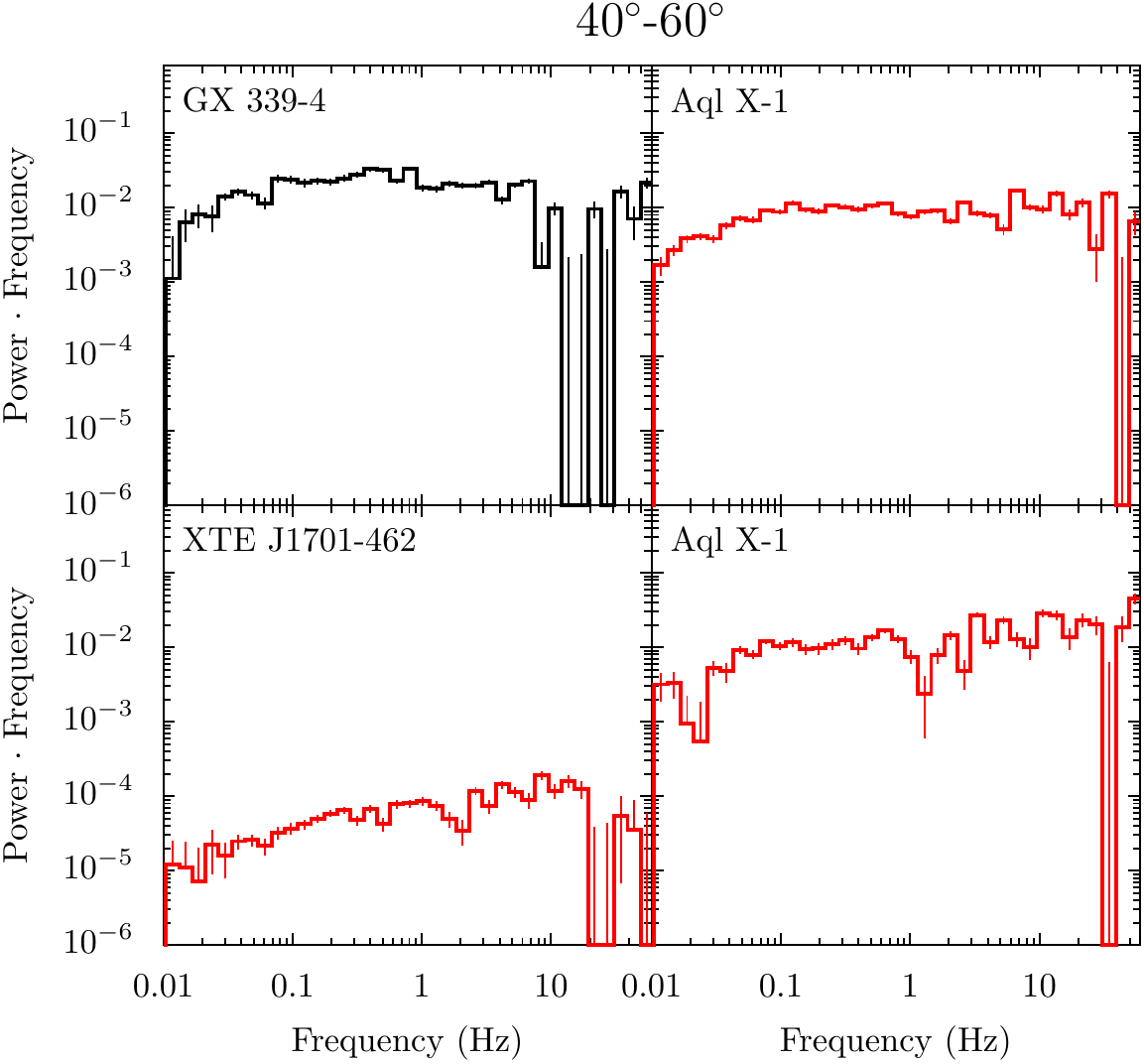}
 \end{subfigure}%
 \begin{subfigure}{0.47\textwidth}
  \includegraphics[width=\textwidth]{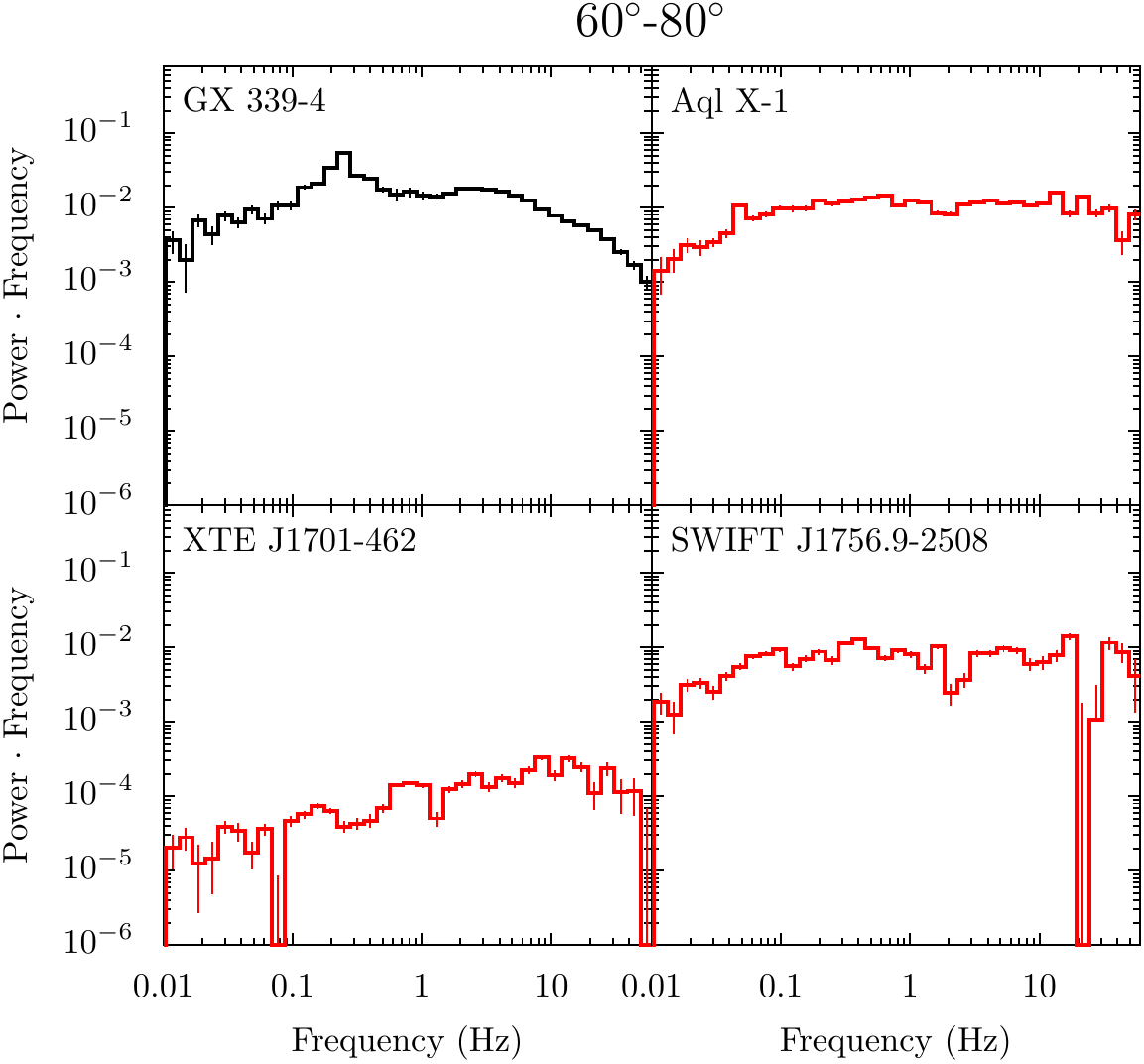}
 \end{subfigure}
 \begin{subfigure}{0.47\textwidth}
  \includegraphics[width=\textwidth]{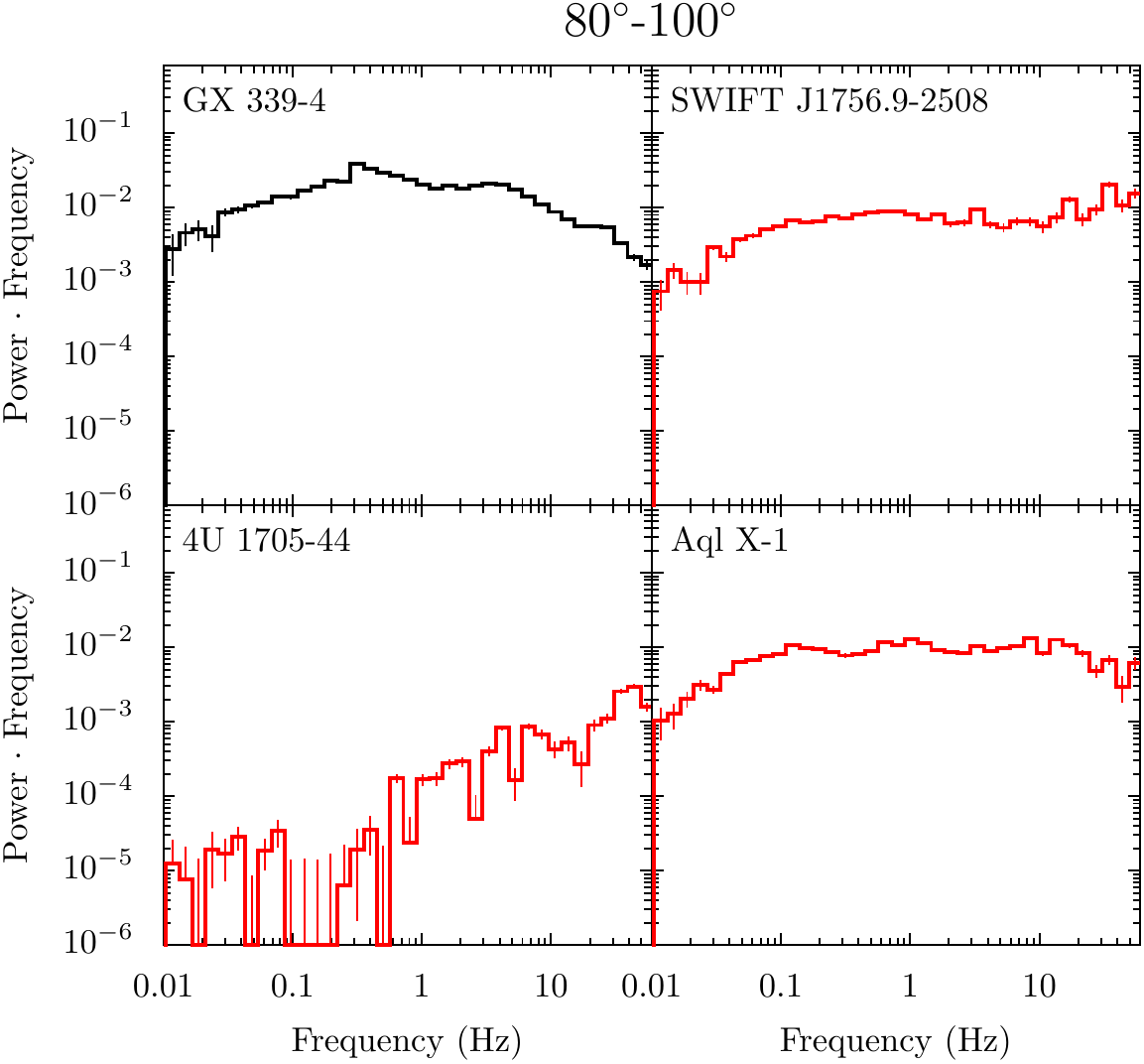}
 \end{subfigure}%
 \begin{subfigure}{0.47\textwidth}
  \includegraphics[width=\textwidth]{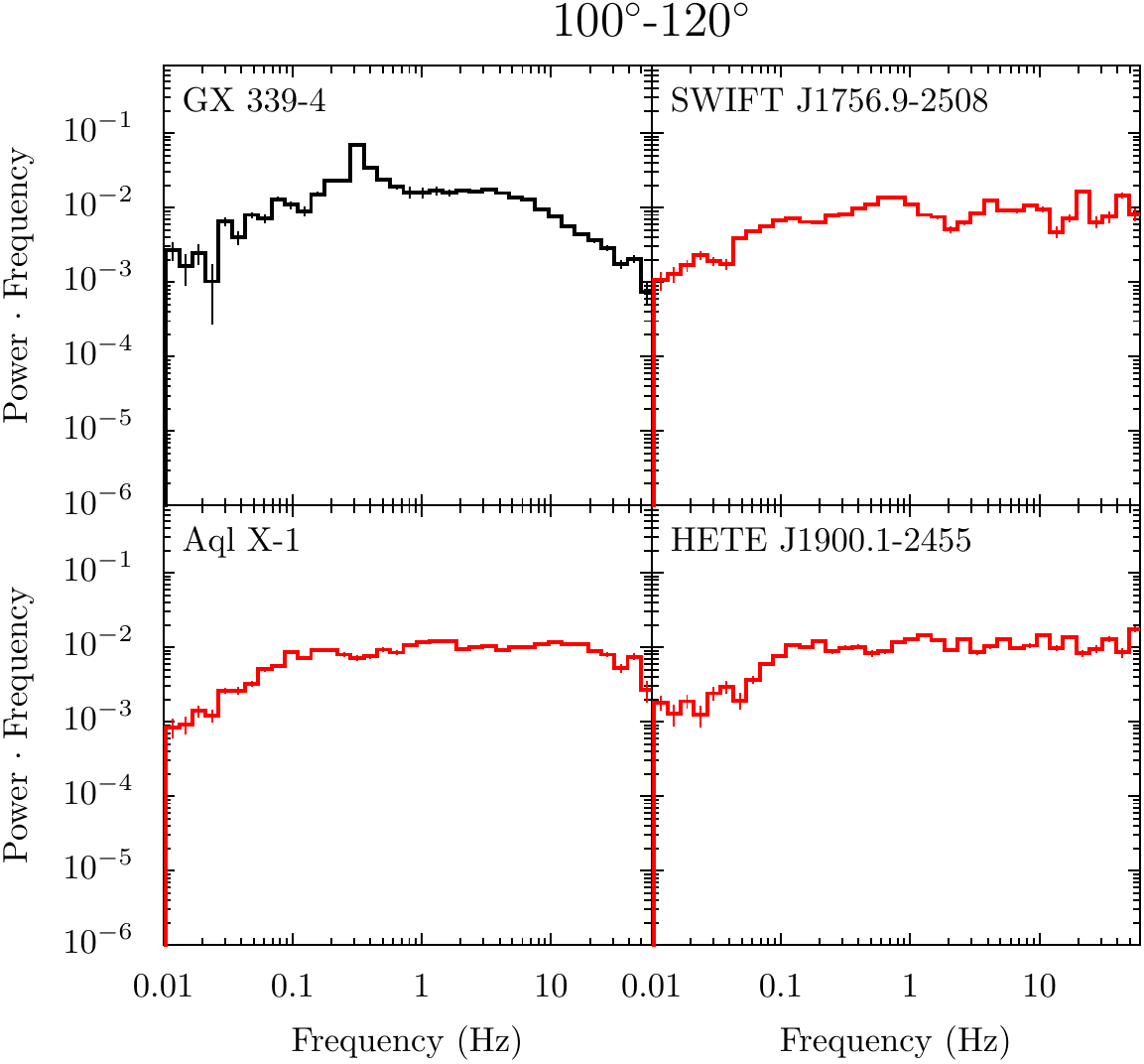}
 \end{subfigure}
 \caption{Representative power spectra within a hue range of 0$^\circ$--120$^\circ$}
 \label{fig:ps_0_120}
\end{figure*}

\begin{figure*}
 \centering
 \begin{subfigure}{0.47\textwidth}
  \includegraphics[width=\textwidth]{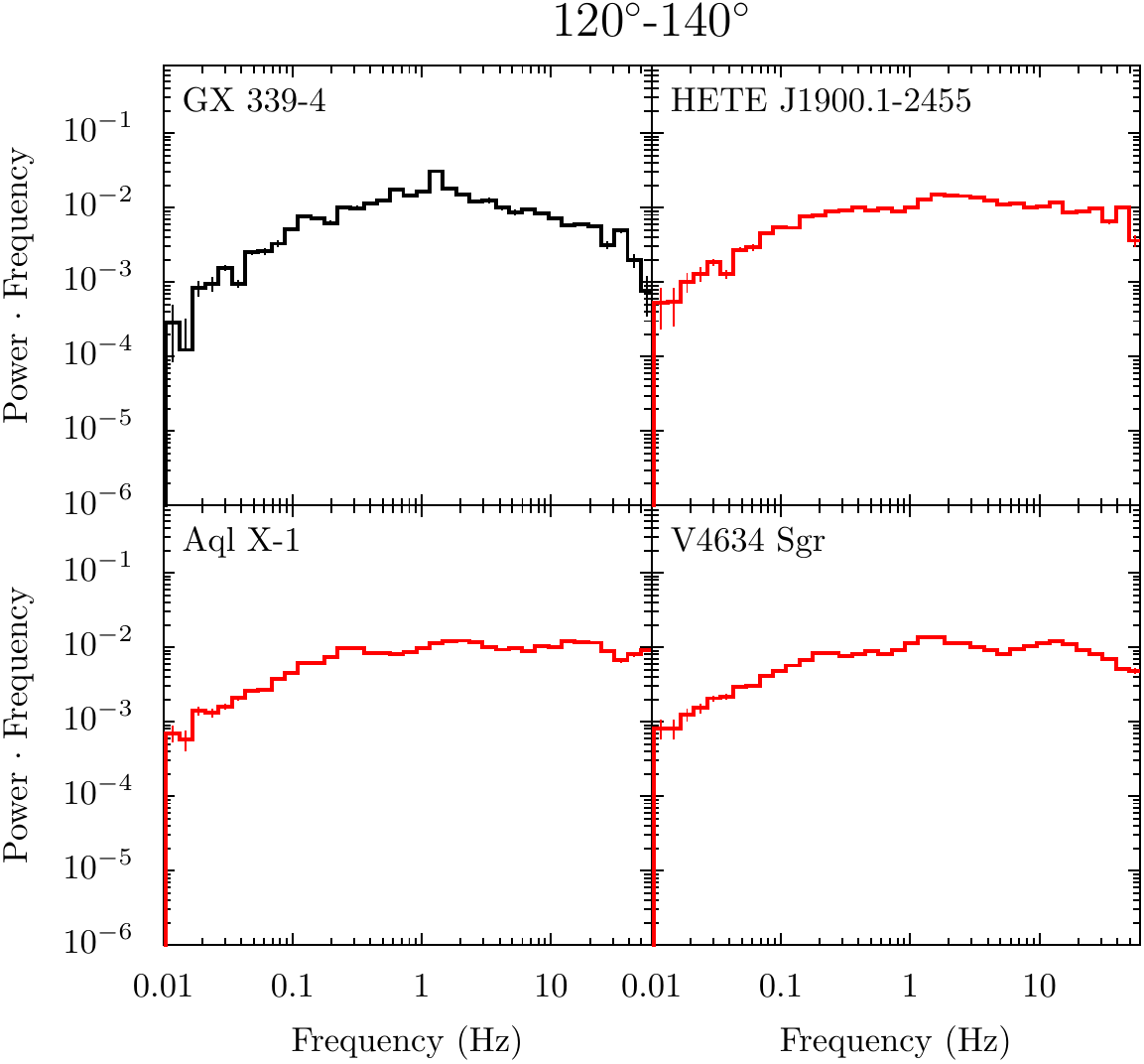}
 \end{subfigure}%
 \begin{subfigure}{0.47\textwidth}
  \includegraphics[width=\textwidth]{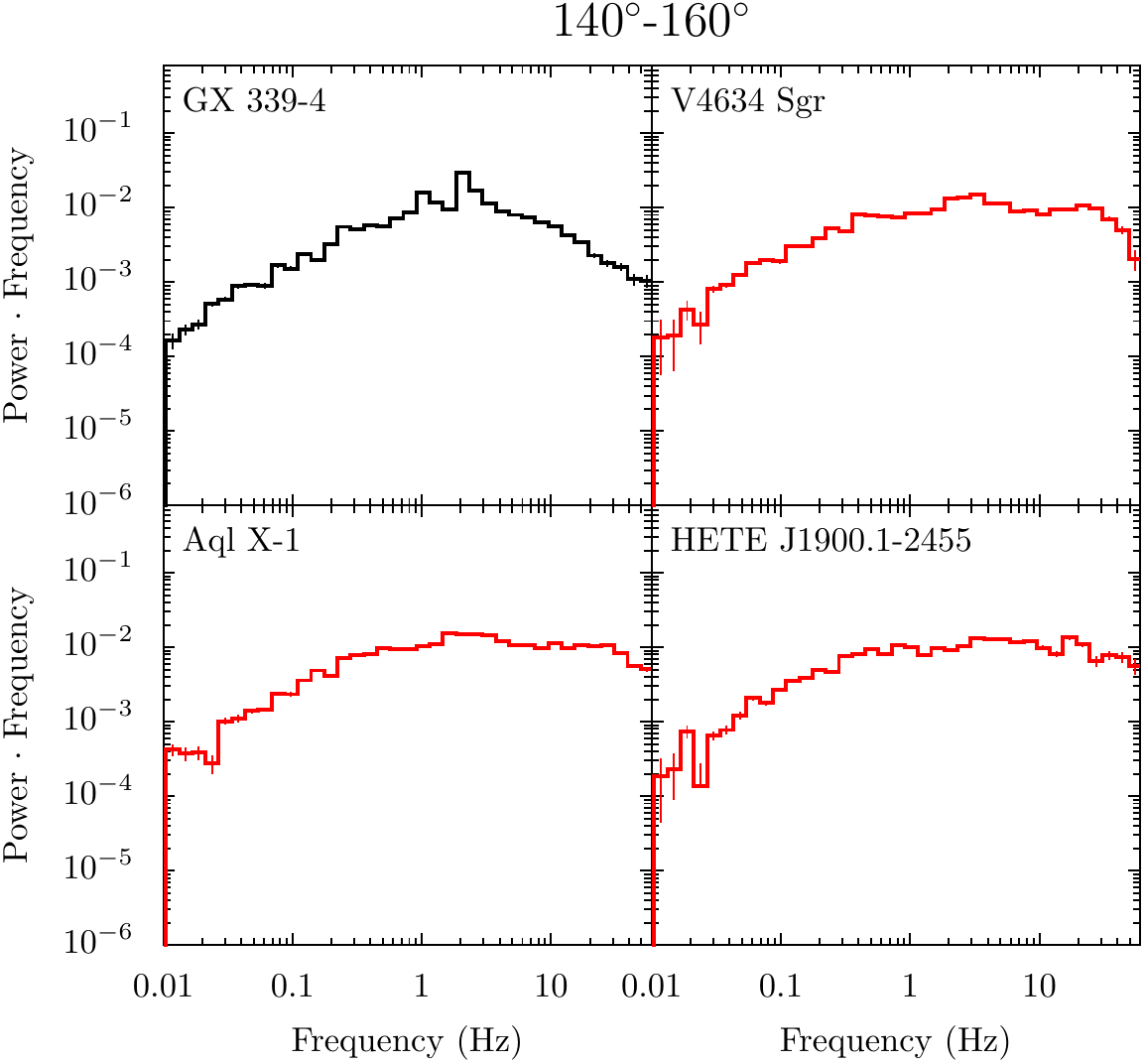}
 \end{subfigure}
 \begin{subfigure}{0.47\textwidth}
  \includegraphics[width=\textwidth]{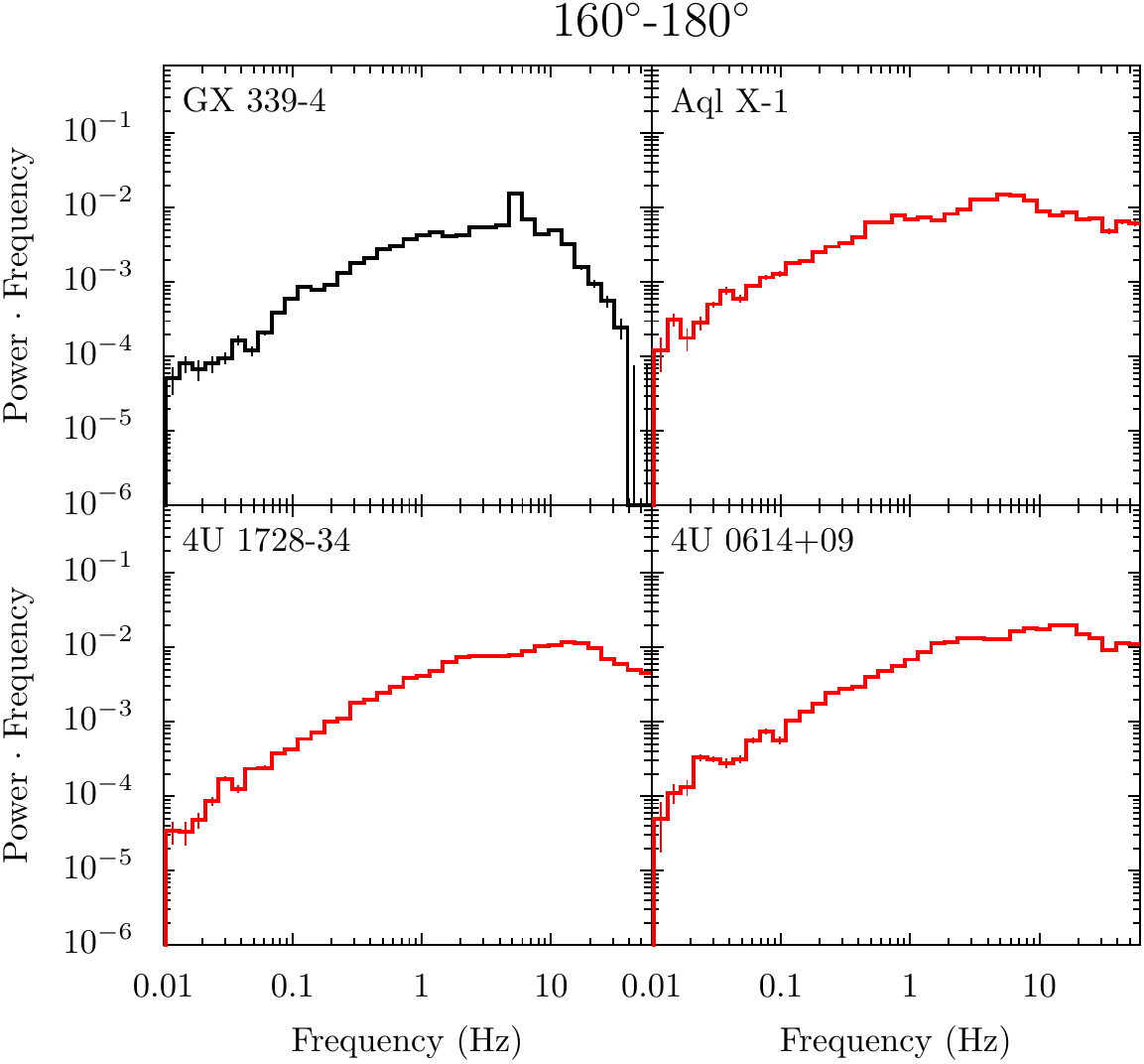}
 \end{subfigure}%
 \begin{subfigure}{0.47\textwidth}
  \includegraphics[width=\textwidth]{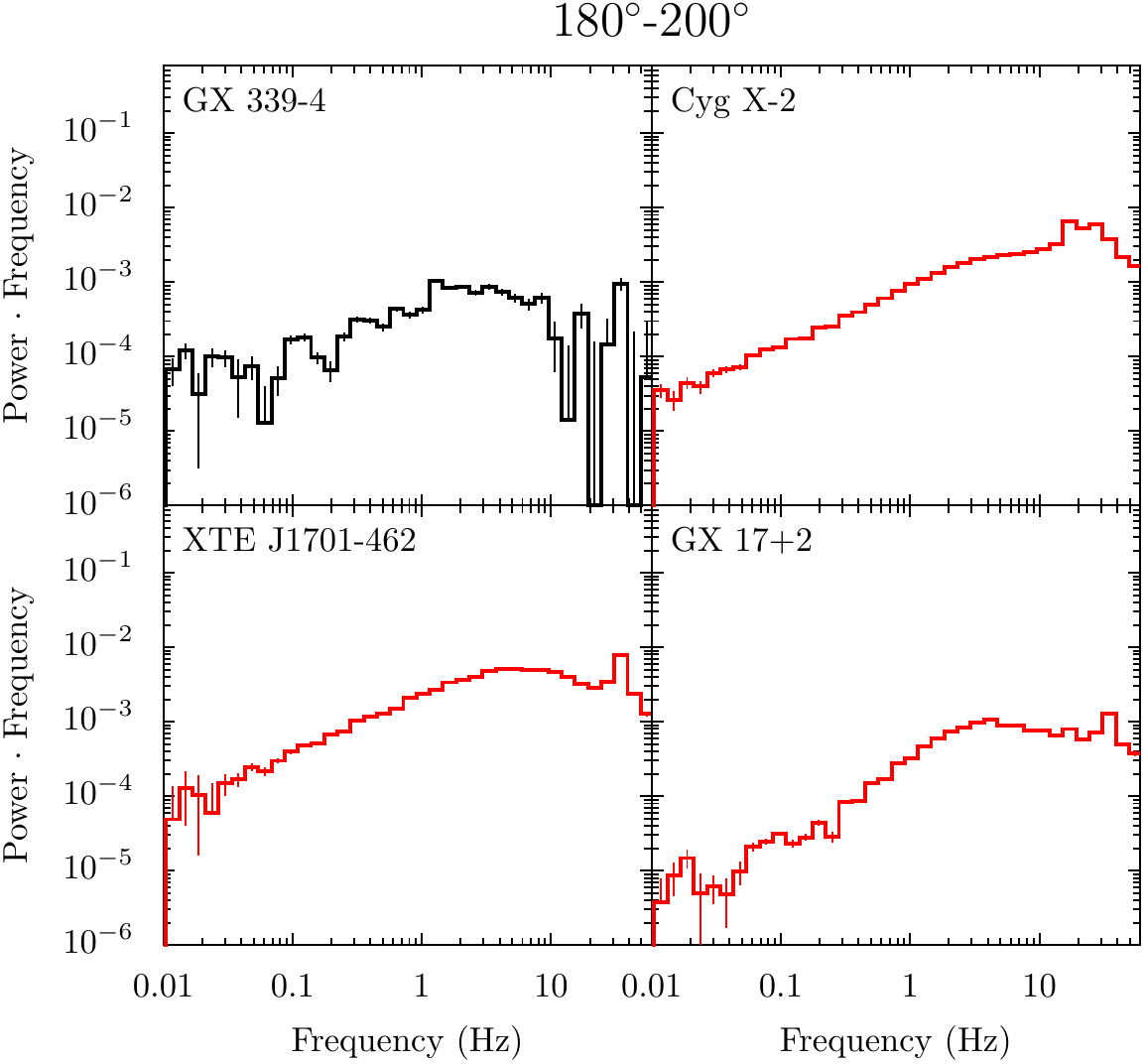}
 \end{subfigure}
 \begin{subfigure}{0.47\textwidth}
  \includegraphics[width=\textwidth]{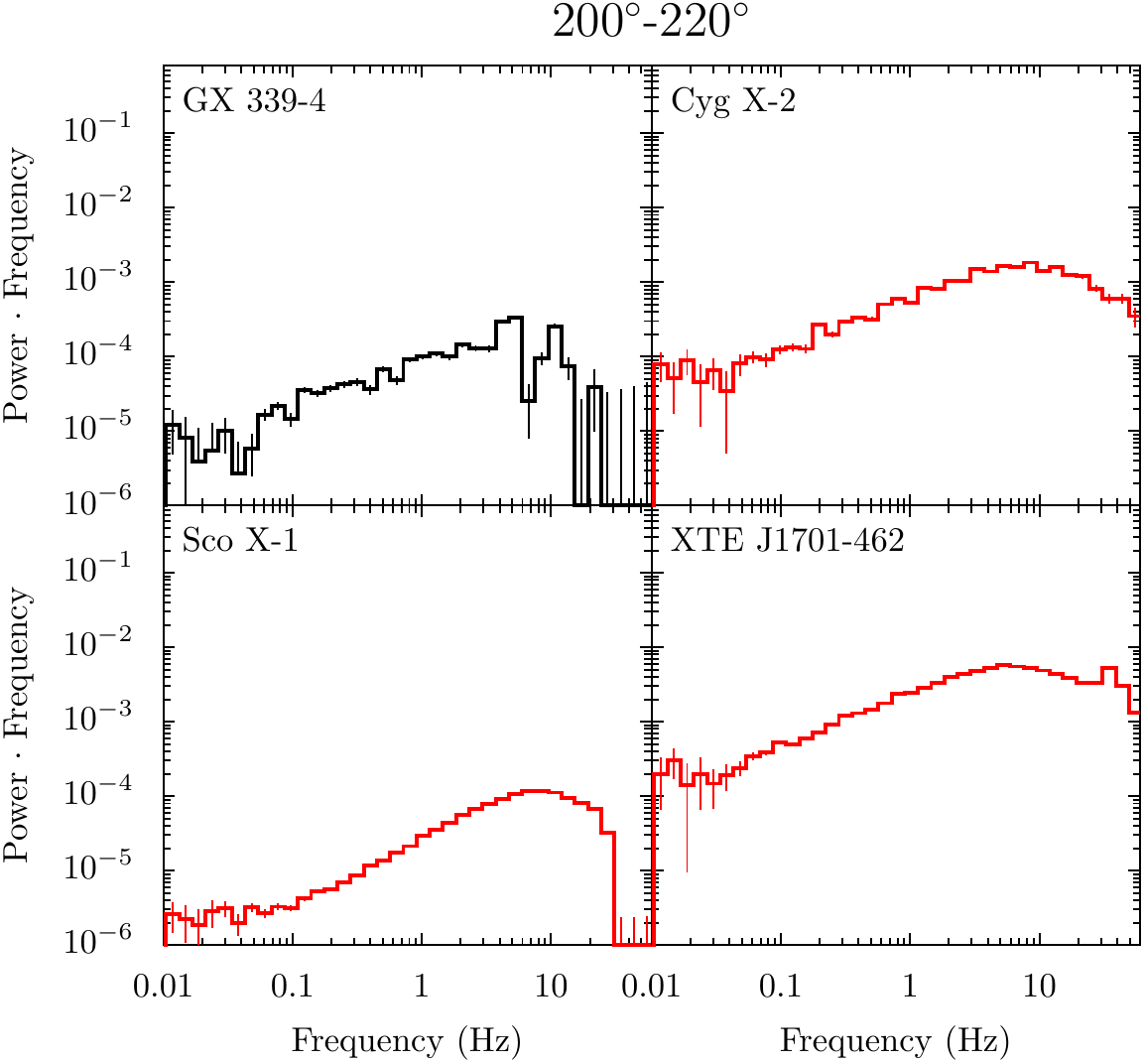}
 \end{subfigure}%
 \begin{subfigure}{0.47\textwidth}
  \includegraphics[width=\textwidth]{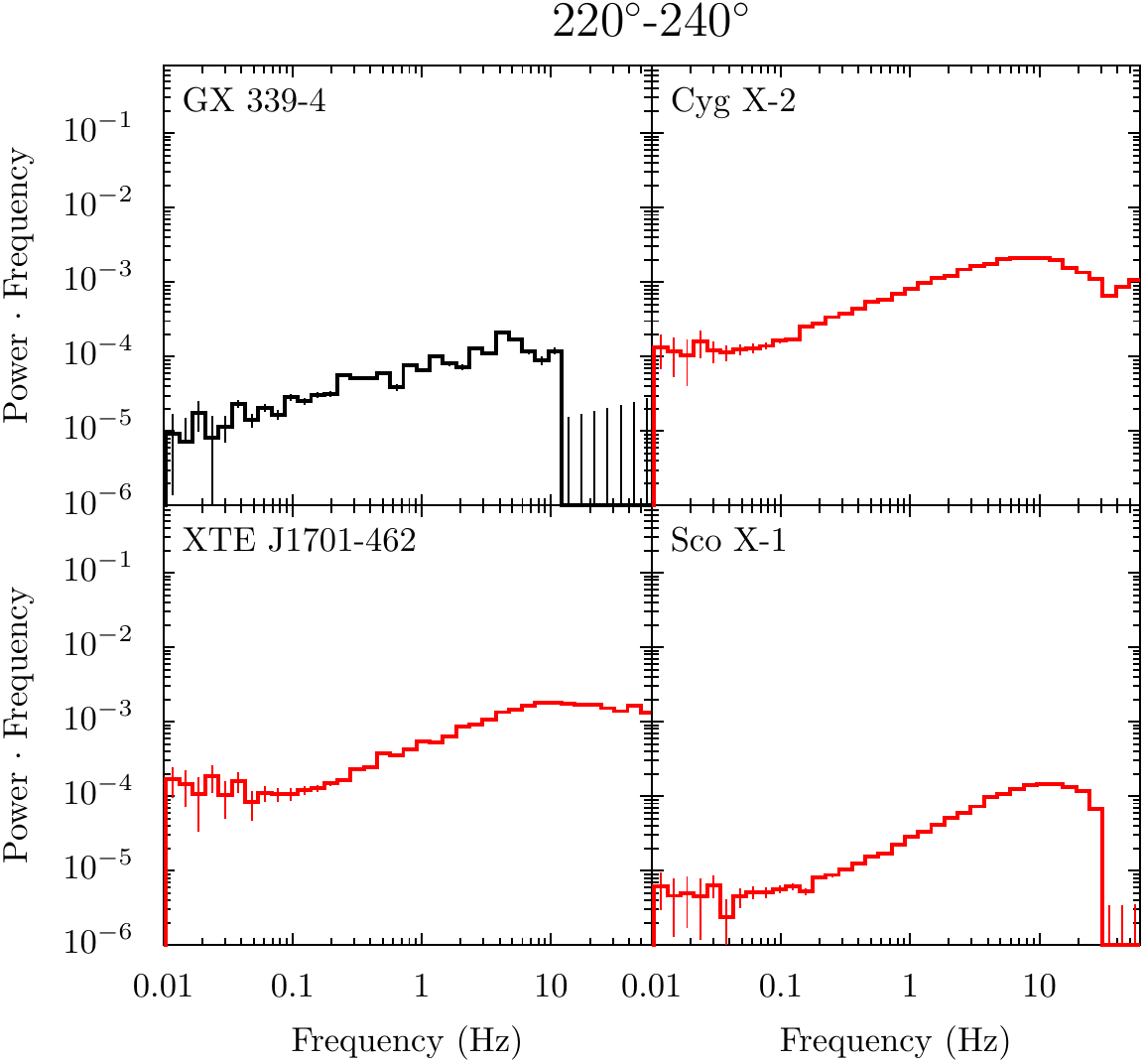}
 \end{subfigure}
 \caption{Representative power spectra within a hue range of 120$^\circ$--240$^\circ$}
 \label{fig:ps_120_240}
\end{figure*}

\begin{figure*}
 \centering
 \begin{subfigure}{0.47\textwidth}
  \includegraphics[width=\textwidth]{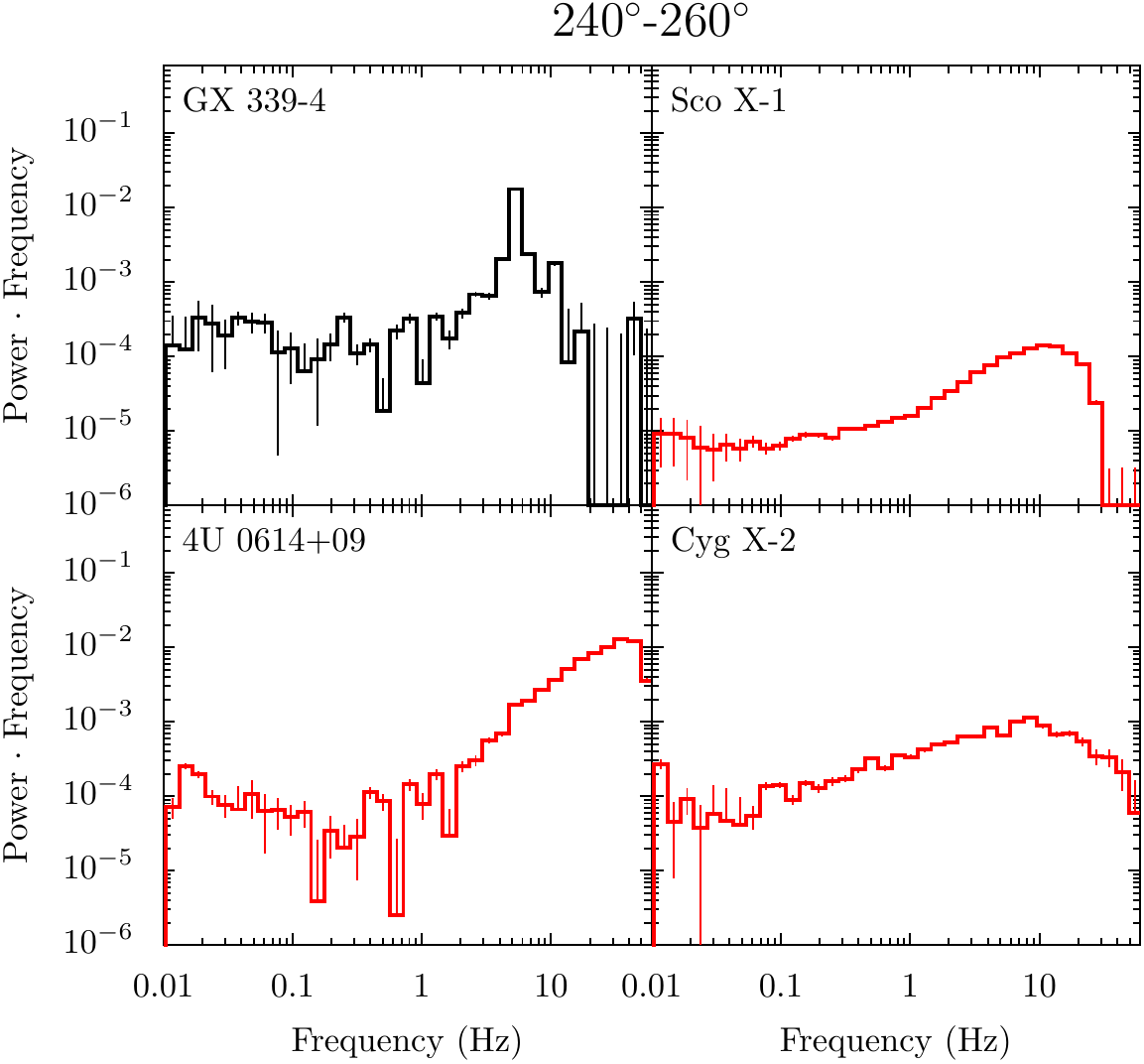}
 \end{subfigure}%
 \begin{subfigure}{0.47\textwidth}
  \includegraphics[width=\textwidth]{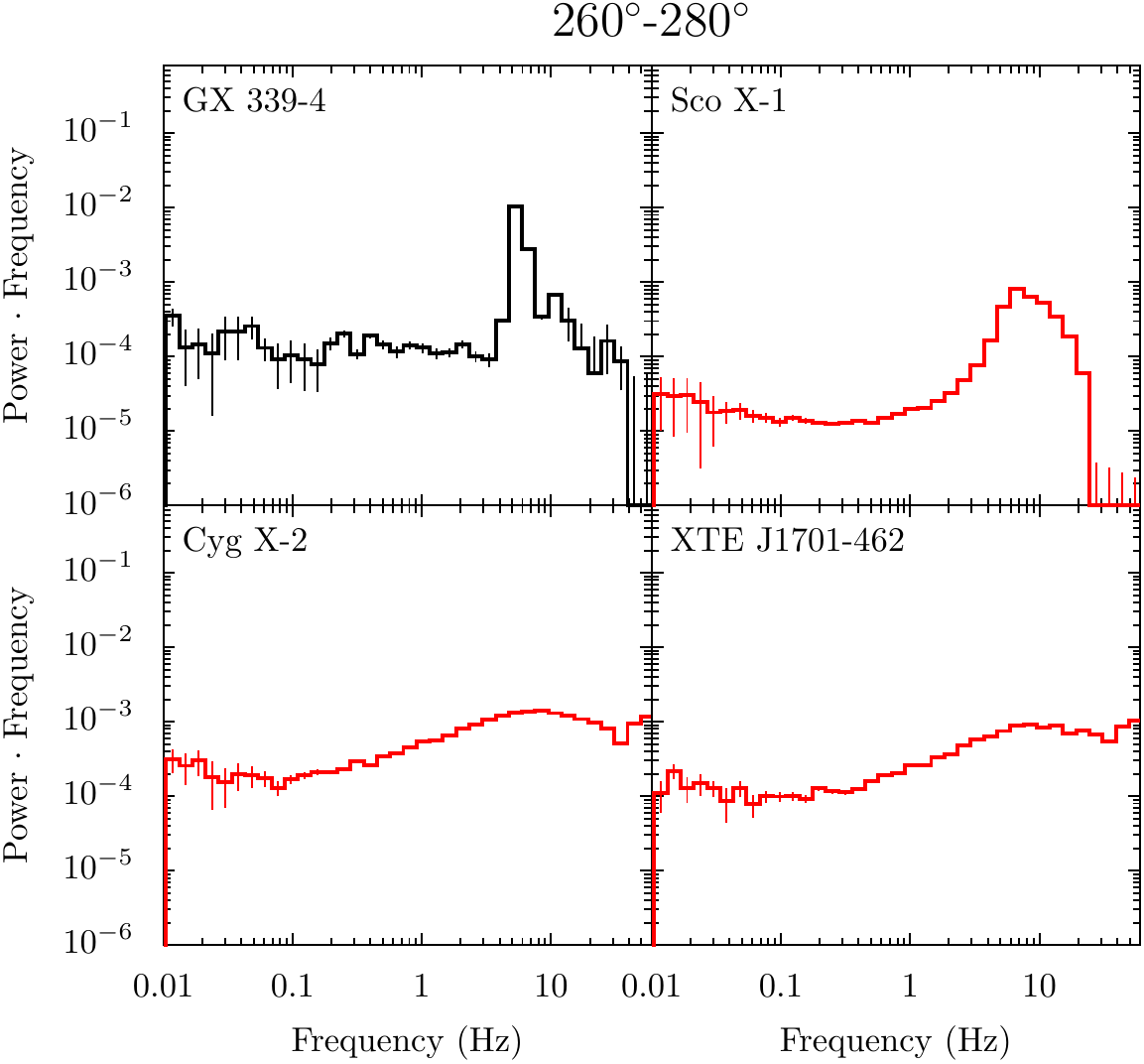}
 \end{subfigure}
 \begin{subfigure}{0.47\textwidth}
  \includegraphics[width=\textwidth]{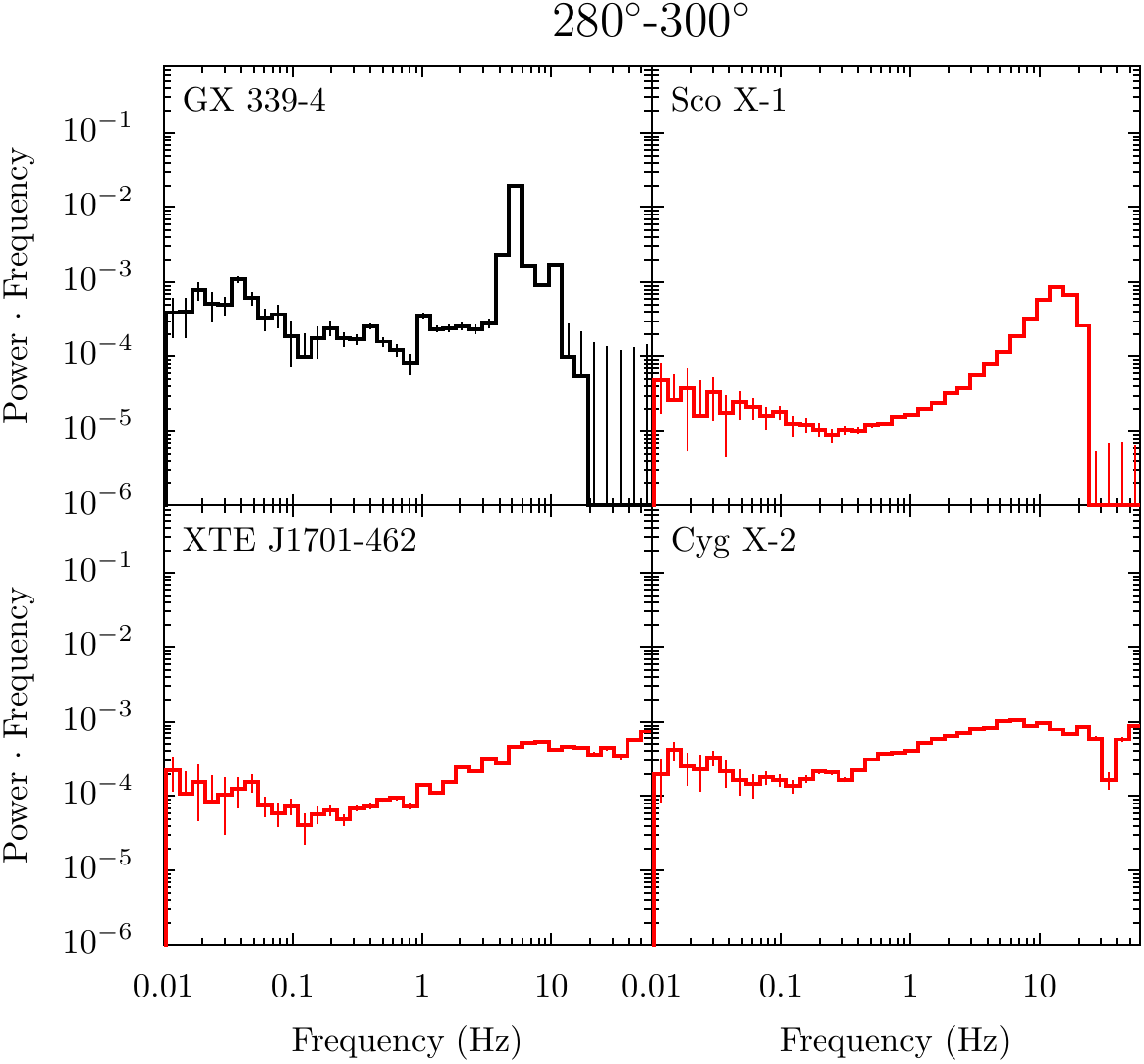}
 \end{subfigure}%
 \begin{subfigure}{0.47\textwidth}
  \includegraphics[width=\textwidth]{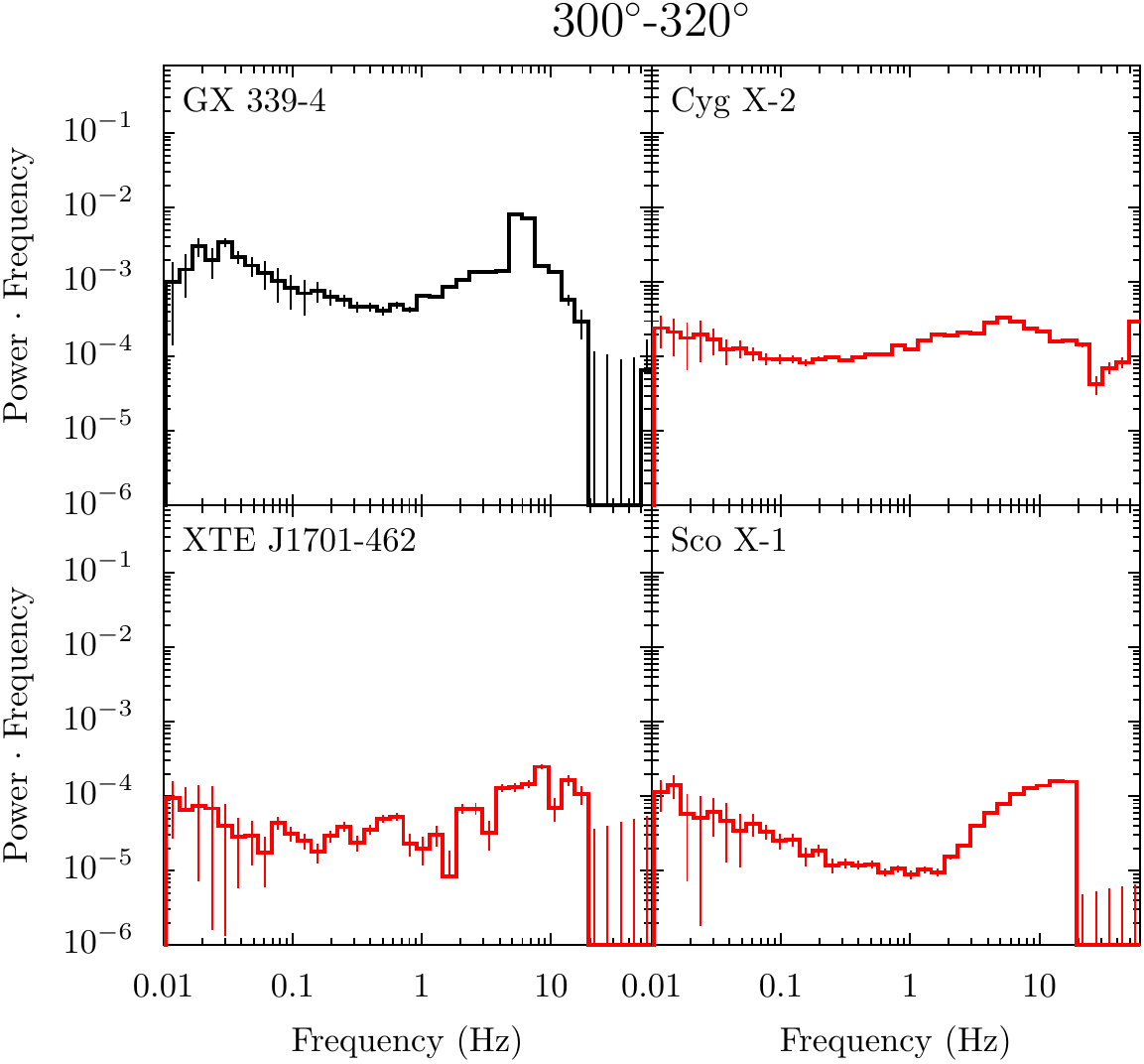}
 \end{subfigure}
 \begin{subfigure}{0.47\textwidth}
  \includegraphics[width=\textwidth]{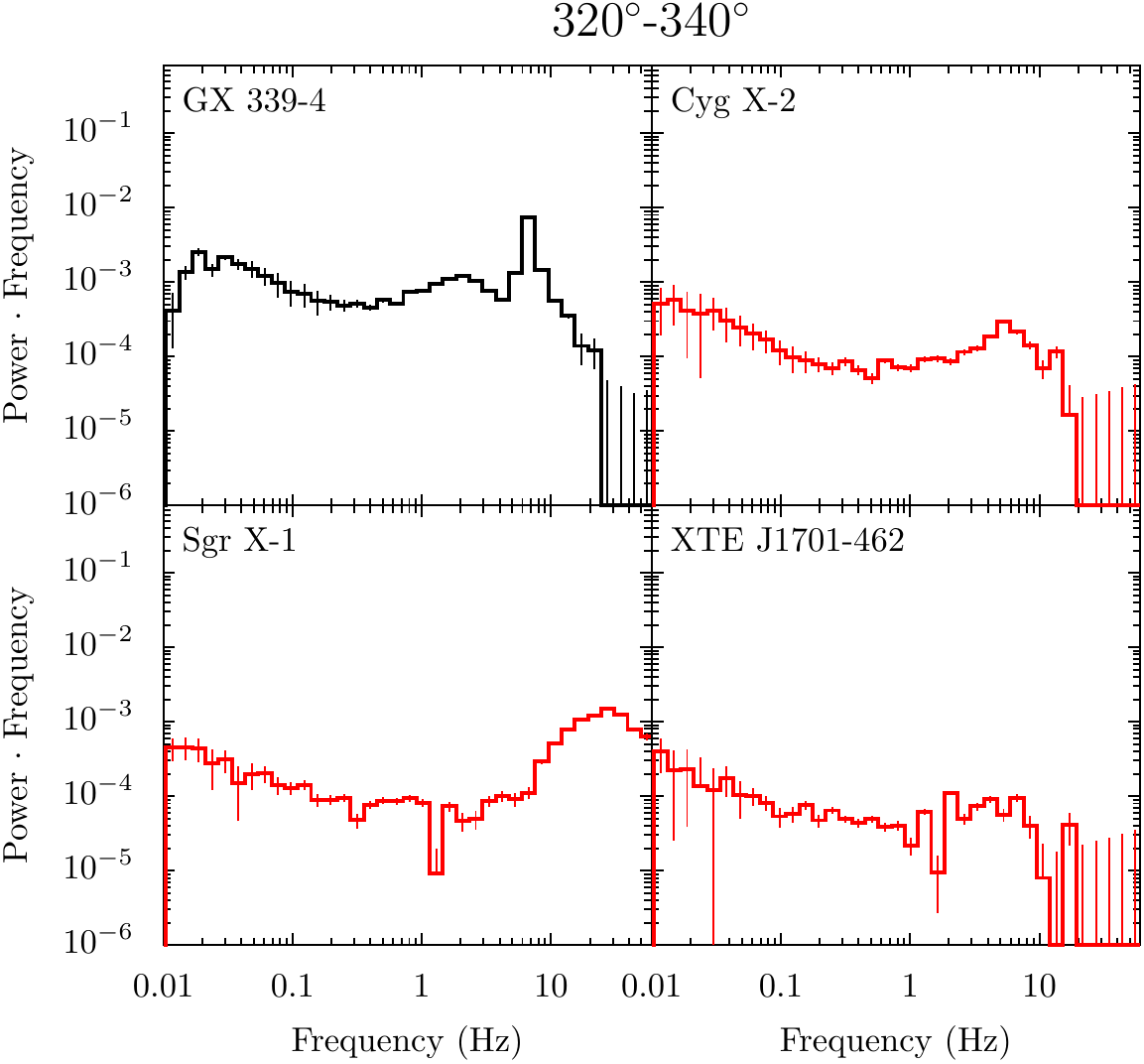}
 \end{subfigure}%
 \begin{subfigure}{0.47\textwidth}
  \includegraphics[width=\textwidth]{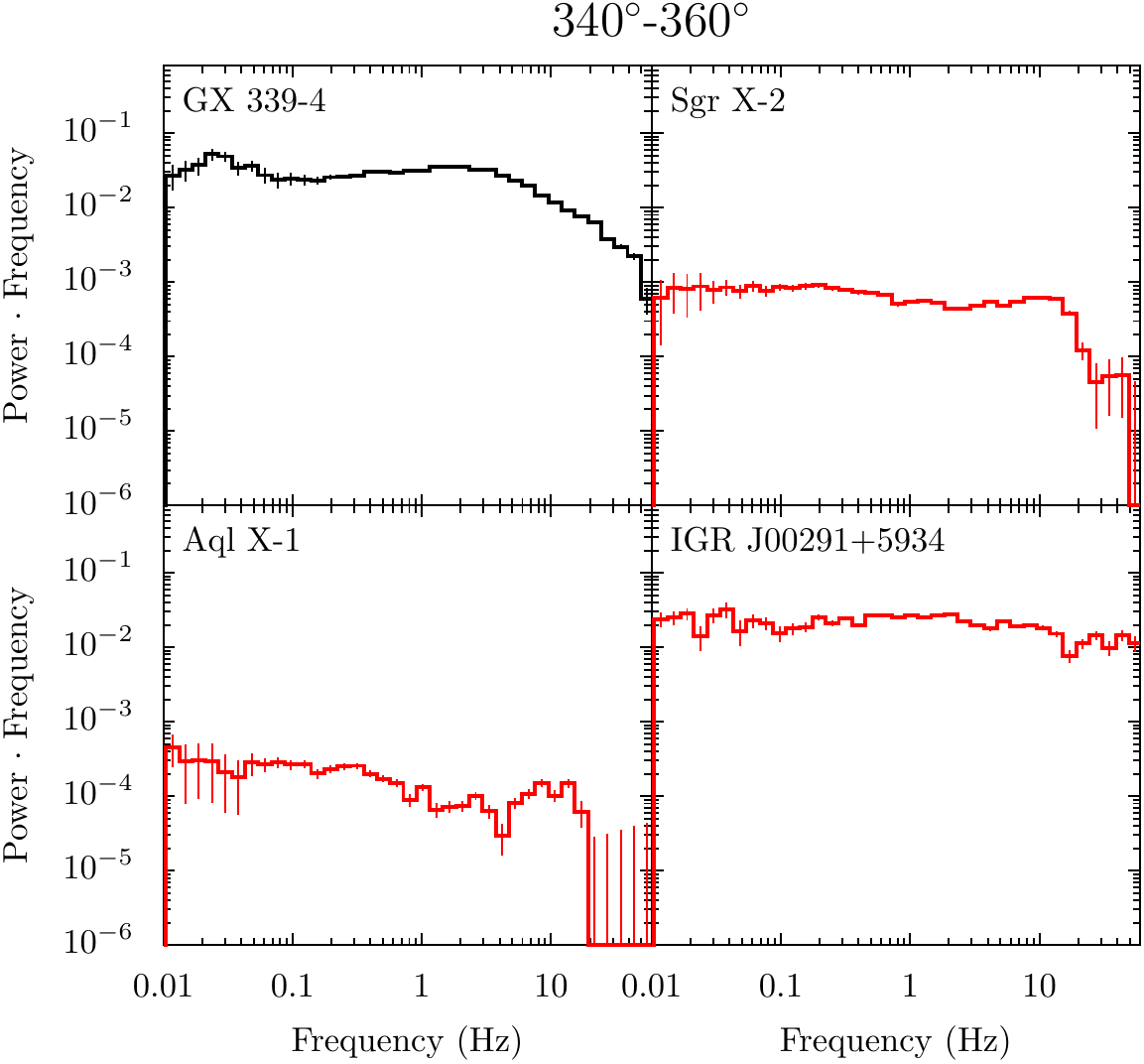}
 \end{subfigure}
 \caption{Representative power spectra within a hue range of 240$^\circ$--360$^\circ$}
 \label{fig:ps_240_360}
\end{figure*}

\section{Power Colour-Colour Diagrams}
\label{sec:pcc}

\begin{figure*}
 \includegraphics[width=\textwidth]{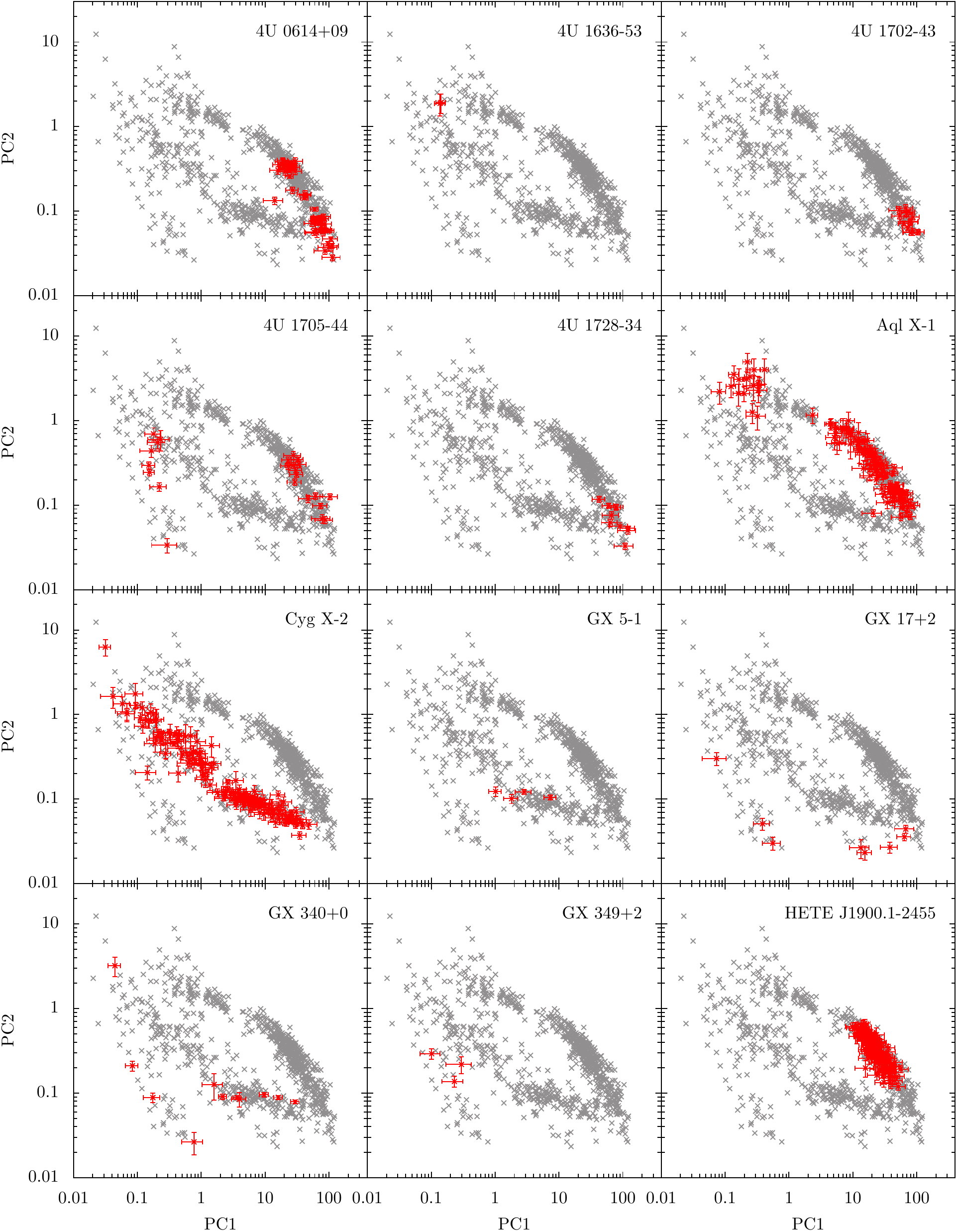}
 \caption{PCC diagrams ranging from 4U to HETE}
 \label{fig:pc_pane_1}
\end{figure*}

\begin{figure*}
 \includegraphics[width=\textwidth]{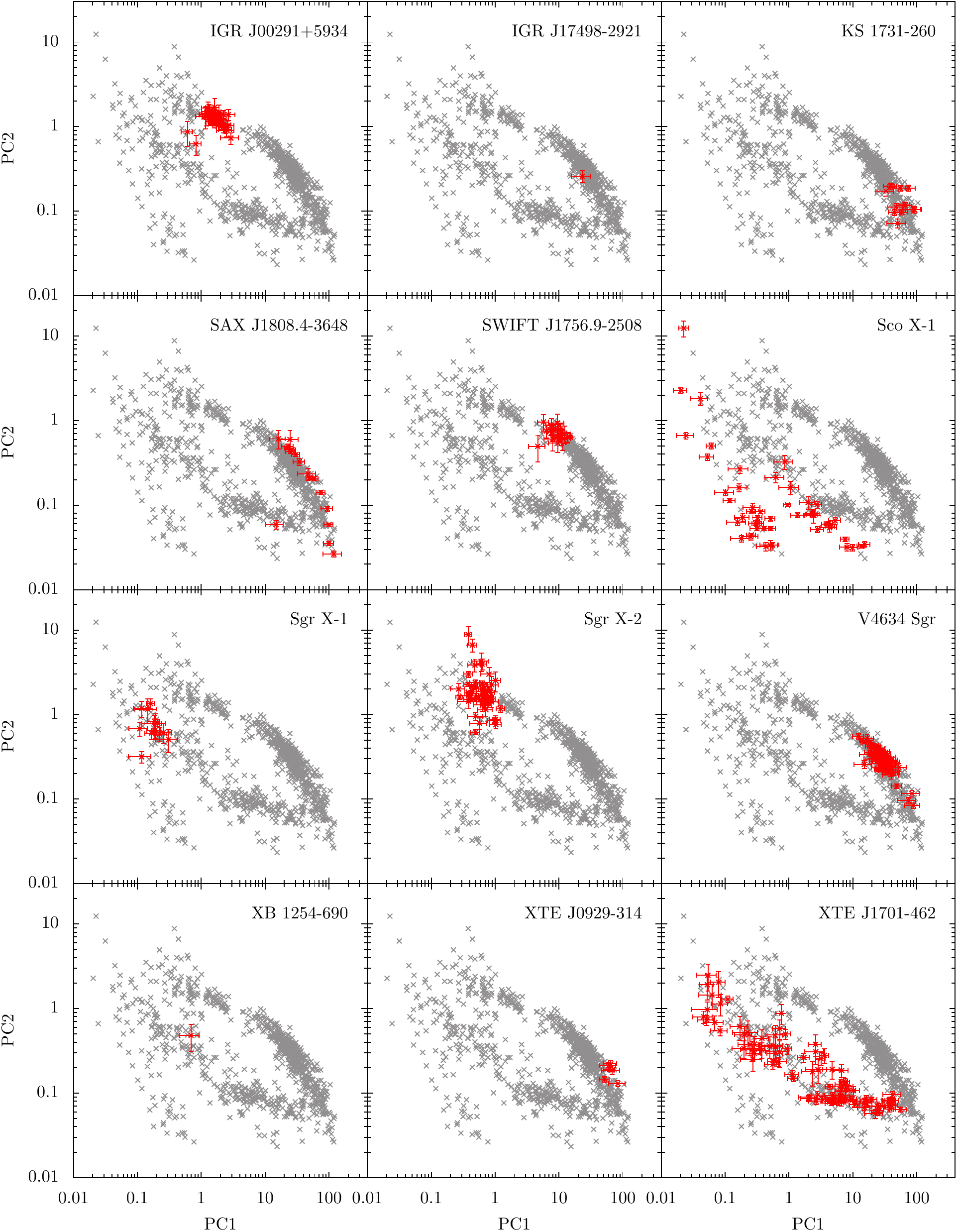}
 \caption{PCC diagrams ranging from IGR to XTE}
 \label{fig:pc_pane_2}
\end{figure*}

\begin{figure*}
 \includegraphics[width=\textwidth]{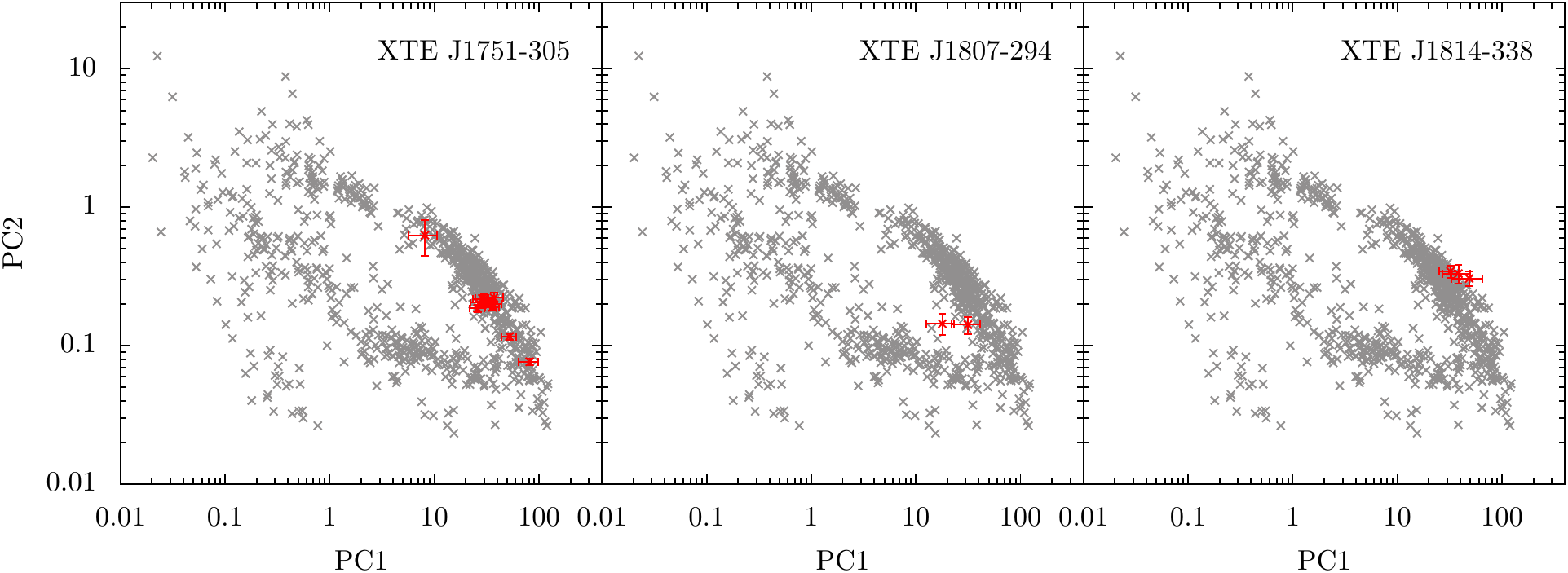}
 \caption{PCC diagrams ranging from XTE to XTE}
 \label{fig:pc_pane_3}
\end{figure*}

\begin{figure*}
 \includegraphics[width=\textwidth]{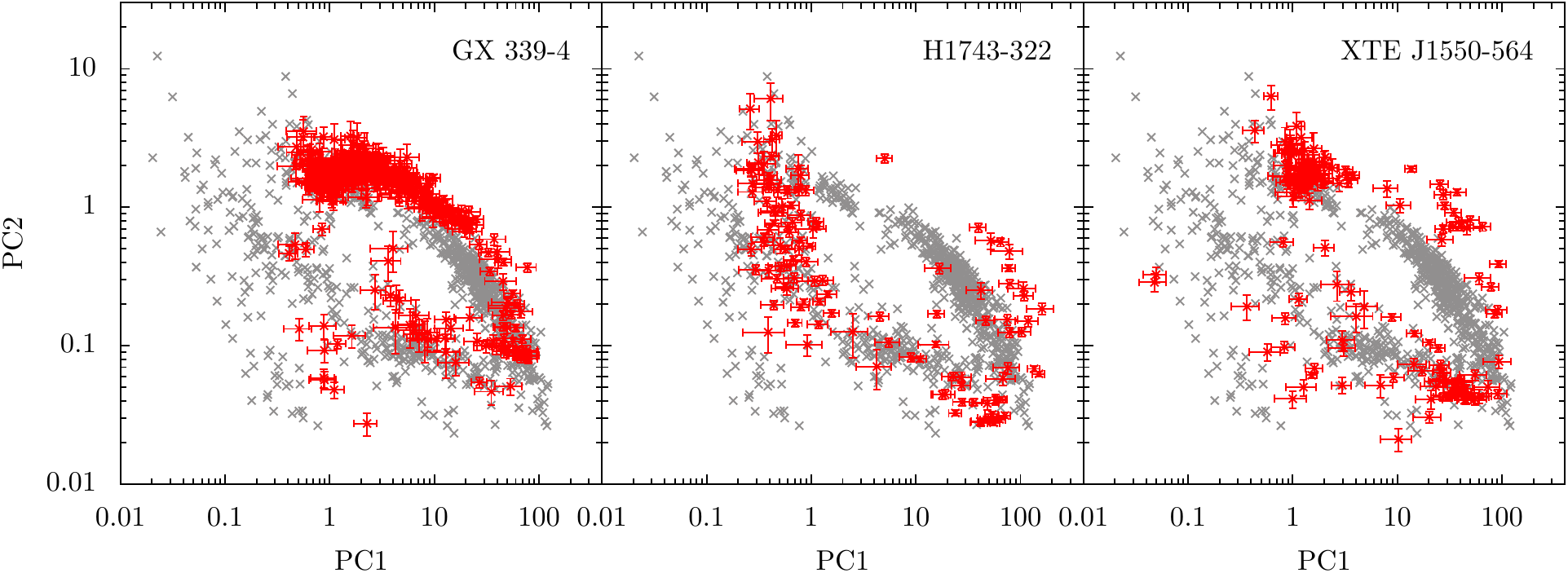}
 \caption{PCC diagrams for black hole LMXBs}
 \label{fig:pc_pane_4}
\end{figure*}

\section{Hardness-Hue Diagrams}
\label{sec:hh}

\begin{figure*}
 \includegraphics[width=\textwidth]{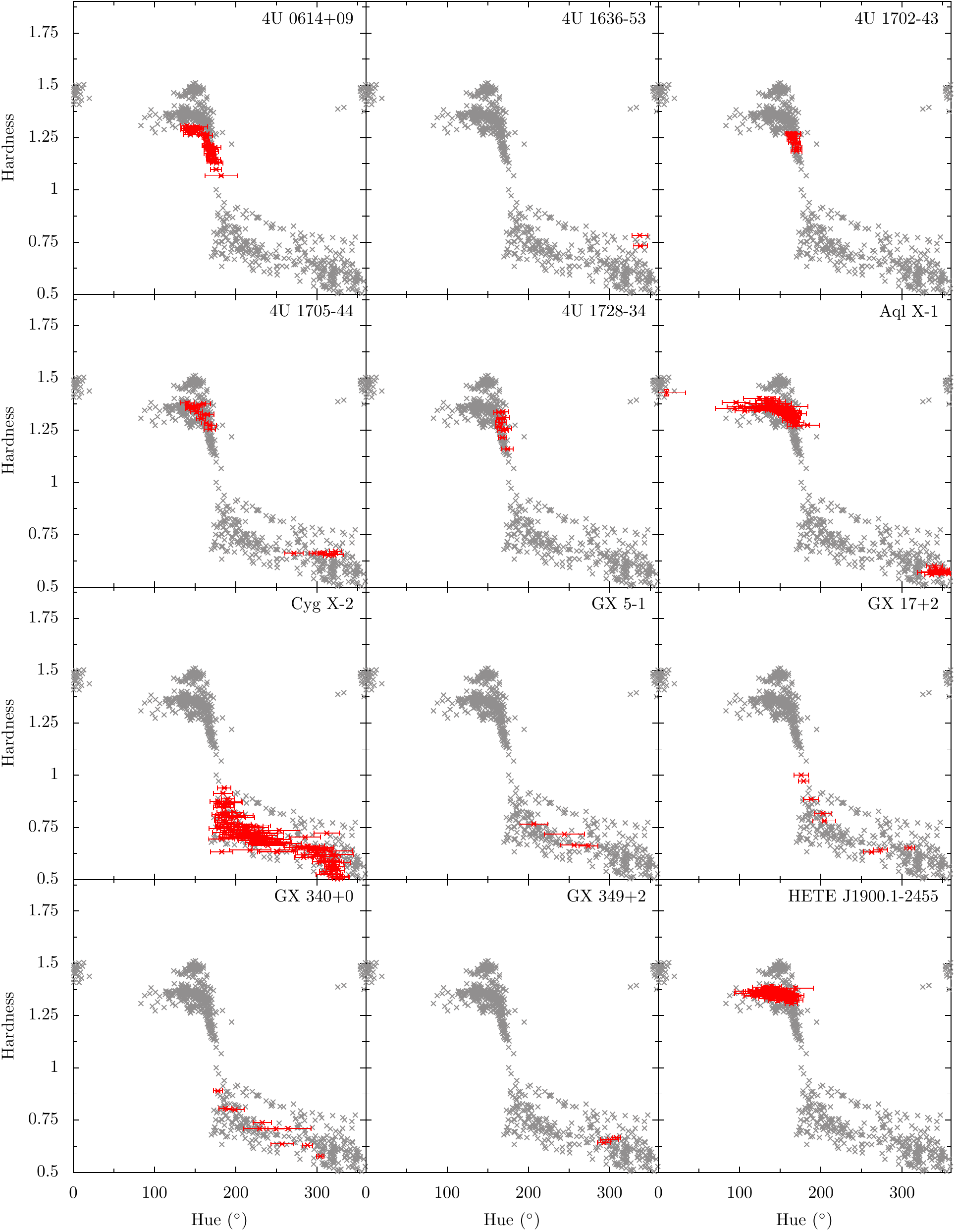}
 \caption{HH diagrams ranging from 4U to HETE}
 \label{fig:hh_pane_1}
\end{figure*}

\begin{figure*}
 \includegraphics[width=\textwidth]{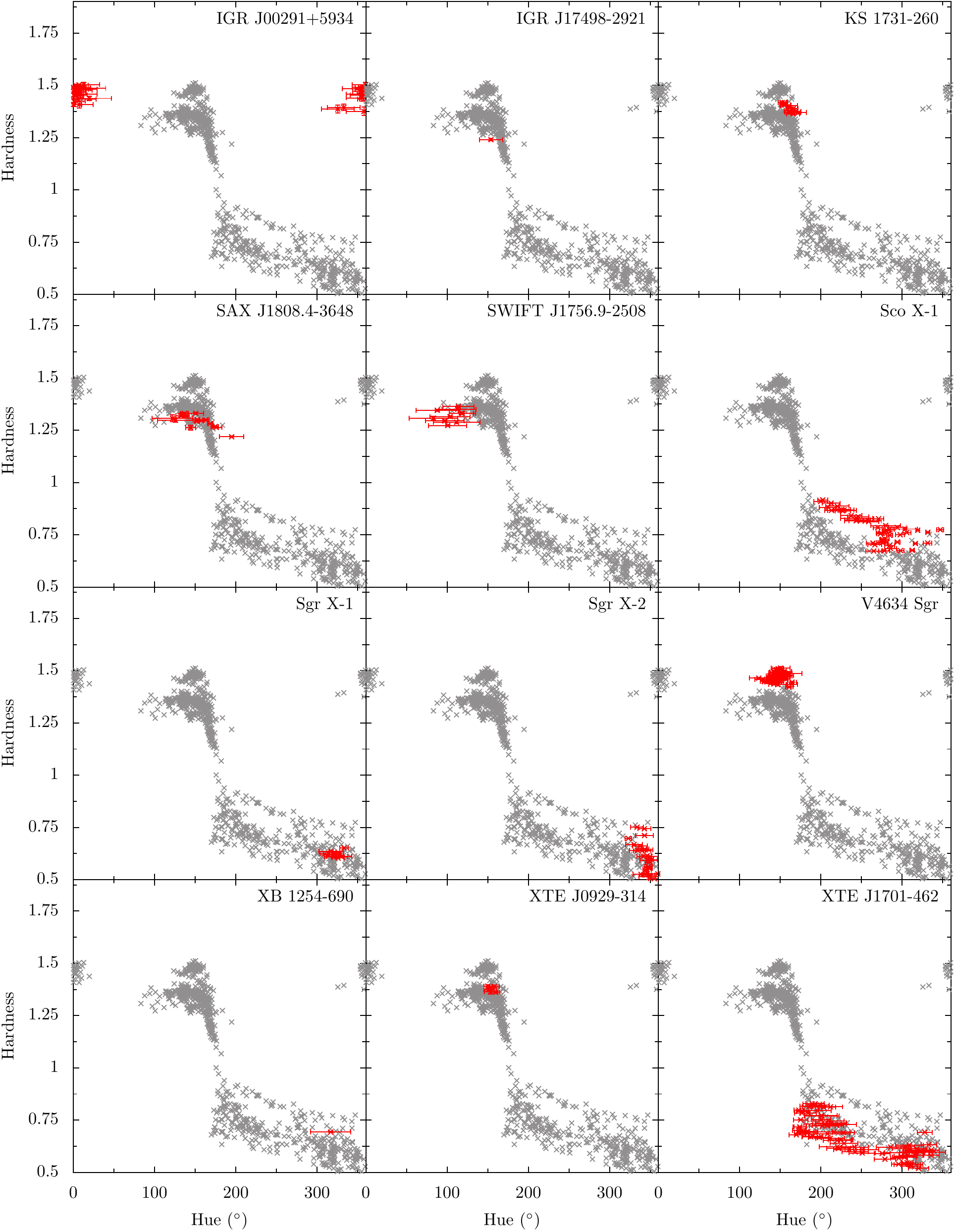}
 \caption{HH diagrams ranging from IGR to XTE}
 \label{fig:hh_pane_2}
\end{figure*}

\begin{figure*}
 \includegraphics[width=\textwidth]{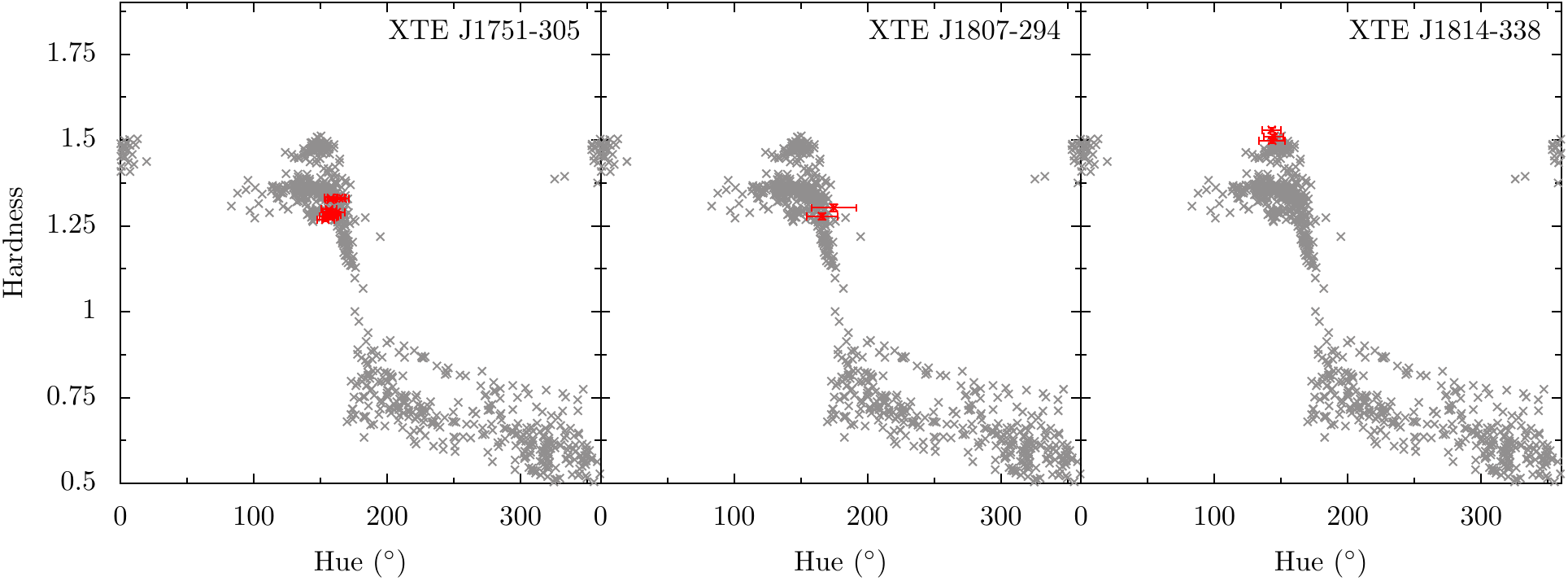}
 \caption{HH diagrams ranging from XTE to XTE}
 \label{fig:hh_pane_3}
\end{figure*}

\begin{figure*}
 \includegraphics[width=\textwidth]{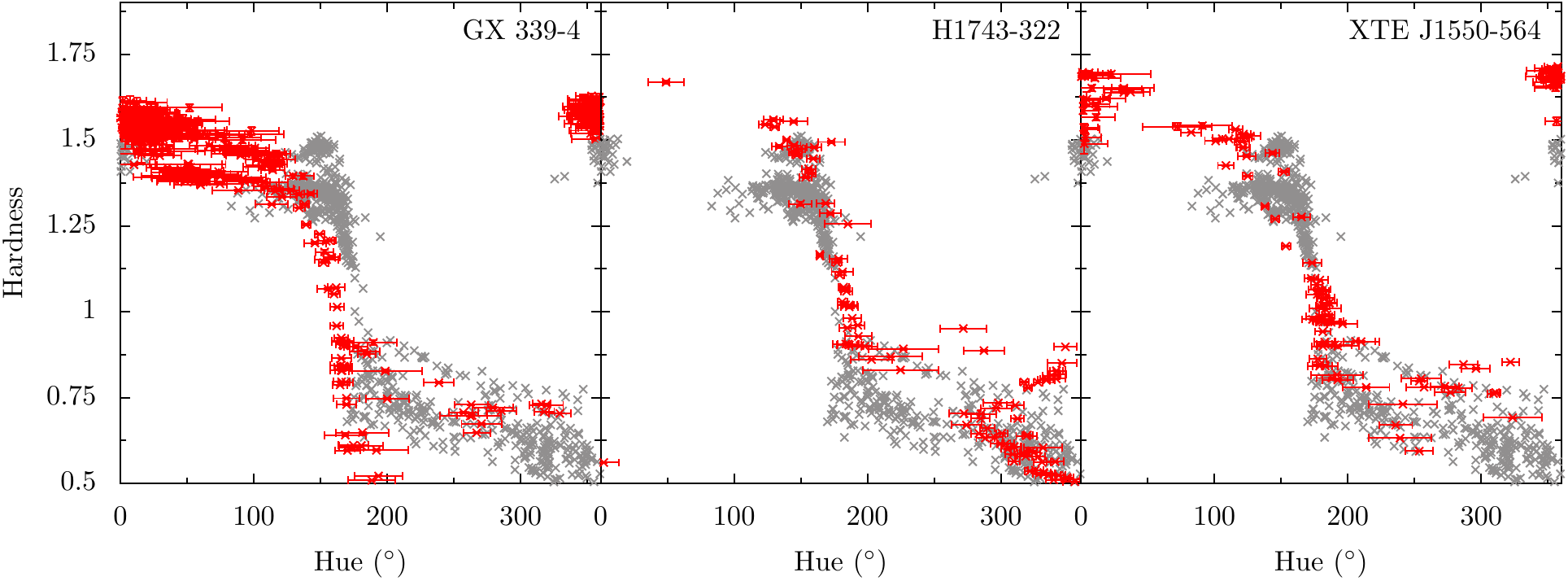}
 \caption{HH diagrams for black hole LMXBs}
 \label{fig:hh_pane_4}
\end{figure*}

\bsp	
\label{lastpage}
\end{document}